\documentclass[a4paper,11pt]{article}
\pdfoutput=1

\usepackage{jcappub} 
\usepackage{amsfonts}
\usepackage[utf8]{inputenc}
\usepackage{graphicx}
\usepackage{dcolumn}
\usepackage{bm}
\usepackage{amsmath}
\DeclareMathOperator{\arcsinh}{arcsinh}
\usepackage{lipsum}
\usepackage{xcolor}
\usepackage{calc}
\usepackage{accents}
\usepackage{comment}
\usepackage{float}

\newcommand{\Planck}{{\sc Planck}}
\newcommand{\BAO}{{\sc BAO}}

\newcommand{\Pantheon}{{\sc Pantheon}}

\newcommand{\As}{\mathcal{A}_{\star}}
\newcommand{\ks}{k_{\star}}

\newcommand{\reffig}[1]{figure~\ref{#1}}

\newcommand{\refsec}[1]{section~\ref{#1}}
\newcommand{\refequ}[1]{eq.~(\ref{#1})}
\newcommand{\refapp}[1]{appendix~\ref{#1}}

\newcommand{\commentout}[1]{} 

\begin{document}
\subheader{ULB-TH/23-04, ...}

\title{Minihalos as probes of  the inflationary spectrum: 
accurate boost factor calculation  and new CMB constraints}

\author[a]{Guillermo Franco Abell\'{a}n}
\emailAdd{g.francoabellan@uva.nl}
\affiliation[a]{ GRAPPA Institute, Institute for Theoretical Physics Amsterdam, \\University of Amsterdam, Science Park 904, 1098 XH Amsterdam, The Netherlands}

\author[b]{and Ga\'{e}tan Facchinetti}
\emailAdd{gaetan.facchinetti@ulb.be}

\affiliation[b]{Service de Physique Th\'eorique, C.P. 225, Universit\'e Libre de Bruxelles,\\ Boulevard du Triomphe, B-1050 Brussels, Belgium}

\date{\today}

\abstract{Although the spectrum of primordial fluctuations has been accurately measured on scales above $\sim 0.1~\rm{Mpc}$, only upper limits exist on smaller scales. In this study, we investigate generic monochromatic enhancements to the $\Lambda$CDM spectrum that trigger the collapse of ultracompact minihalos (UCMHs) well before standard structure formation. We refine previous treatments by considering a mixed population of halos with different density profiles, that should realistically arise as a consequence of late-time accretion and mergers. Assuming that dark matter (DM) can self-annihilate, we find, as expected, that UCMHs can greatly enhance the annihilation rate around recombination, significantly imprinting the cosmic microwave background (CMB) anisotropies. However, we provide additional insight on the theoretical uncertainties that currently impact that boost and which may affect late-time probes such as the 21 cm line or $\gamma$-ray signals. We derive constraints on the primordial power spectrum on small scales using the {\tt ExoCLASS/HYREC} codes and the \Planck~legacy data. We account for the velocity dependence of the DM annihilation cross-section ($s$- or $p$-wave), annihilation channel, the DM particle mass and the inclusion of late-time halo mergers. Our $s$-wave constraints are competitive with previous literature, excluding primordial amplitudes $\As \gtrsim 10^{-6.5}$ at wavenumbers $k \sim 10^4-10^7 \ \rm{Mpc}^{-1}$. For the first time, we highlight that even $p$-wave processes have constraining power on the primordial spectrum for cross-sections still allowed by currently the strongest astrophysical constraints. Finally, we provide an up-to-date compilation of the most stringent limits on the primordial power spectrum across a wide range of scales. 
}

\arxivnumber{2304.02996}

\maketitle

\section{Introduction}

The cold dark matter (CDM) scenario predicts that small structures in the matter-dominated era of the Universe collapse and merge  to form increasingly larger halos, leading to hierarchical structure formation. The distribution of these halos throughout the Universe is determined by the primordial curvature power spectrum $\mathcal{P}_{\mathcal{R}}(k)$, which sets the initial conditions for the evolution of the dark matter (DM) density field. Hence, the primordial power spectrum not only contains valuable information about the early universe but is also crucial for studying the formation and evolution of cosmic structure.

Current data point towards an almost scale-invariant spectrum, as predicted by vanilla slow-roll inflation. In the $\Lambda$CDM paradigm, the dimensionless curvature power spectrum is usually parametrized by
\begin{equation}
    \mathcal{P}_{\rm PL}(k) = \mathcal{A}_{\rm s} \left(\frac{k}{k_0}\right)^{n_{\rm s}-1} \, ,
    \label{eq:LambdaCDMpower}
\end{equation}
where PL stands for \emph{power law}. The amplitude $\mathcal{A}_{\rm s}$ and spectral index $n_{\rm s}$ are determined from CMB anisotropies \cite{Planck:2018vyg} and Lyman-$\alpha$ observations \cite{Bird:2010mp}. The former tightly constrains  $\mathcal{A}_{\rm s}  \simeq 2.1 \times 10^{-9}$ and $n_{\rm s} \simeq 0.96$ (the reference wavenumber being fixed to $k_0 = 0.05$~Mpc$^{-1}$). Nonetheless, the full shape of the power spectrum remains almost unconstrained for modes $k \gtrsim 3$~Mpc$^{-1}$, where it can deviate from the herein above power-law template. For instance, various exotic inflationary scenarios predict enhancements, like single-field \cite{Ivanov:1994pa, Garcia-Bellido:2017mdw, Byrnes:2018txb} or multi-field models with hybrid \cite{Clesse:2015wea, Garcia-Bellido:1996mdl} or double inflation \cite{Silk:1986vc}. Certain exotic thermal histories, such as an early matter dominated era \cite{Erickcek:2011us,Barenboim:2013gya,Fan:2014zua}, an era dominated by a fast rolling scalar field \cite{Redmond:2018xty} or primordial magnetic fields \cite{Ralegankar:2023pyx}, also predict enhancements of small-scale perturbations.

A large amplification to the primordial matter power spectrum of the order $\mathcal{P}_{\mathcal{R}}(k) \sim 10^{-2}$ would result in a significant production of primordial black holes (PBHs)  \cite{Zeldovich:1967lct, Hawking:1971ei, Carr:1974nx}, which could represent a substantial fraction (if not all) of the DM content \cite{Carr:2016drx, Carr:2020xqk}. More moderate enhancements on small scales can also still lead to detectable signals through the production of CMB spectral distortions and gravitational waves \cite{Unal:2020mts}. Spectral observations of the cosmic microwave background by the Far-Infrared Absolute Spectrophotometer (FIRAS) have constrained the power spectrum on modes between $10~{\rm Mpc}^{-1}$ and $10^4~{\rm Mpc}^{-1}$, setting an upper bound of $\mathcal{P}_{\mathcal{R}}(k) \lesssim 10^{-4}$ \cite{Fixsen:1996nj, Chluba:2012we}. Pulsar timing arrays currently provide the strongest bounds on gravitational waves at larger modes, with $\mathcal{P}_{\mathcal{R}}(k) < \mathcal{O}(10^{-3} - 10^{-2})$ for $10^6~{\rm Mpc}^{-1} \lesssim k \lesssim 10^7~{\rm Mpc}^{-1}$ \cite{Inomata:2018epa}. The upcoming Square Kilometer Array (SKA) telescope will further probe the power spectrum on the same modes, with sensitivity to $\mathcal{P}_{\mathcal{R}}(k) > 10^{-5}$ \cite{Moore:2014lga, Inomata:2018epa}. Eventually, the next generation of gravitational interferometers like LISA, BBO or the Einstein Telescope will be sensitive to $\mathcal{P}_{\mathcal{R}}(k) > \mathcal{O}(10^{-6} - 10^{-4})$ at even larger modes, $10^{10}~{\rm Mpc}^{-1} \lesssim k \lesssim 10^{15}~{\rm Mpc}^{-1}$ \cite{Inomata:2018epa}.  For further details on the current and upcoming constraints see Refs.~\cite{Byrnes:2018txb, Green:2020jor} and references within.

\begin{figure}[t!]
\centering
\includegraphics[width=0.99\linewidth]{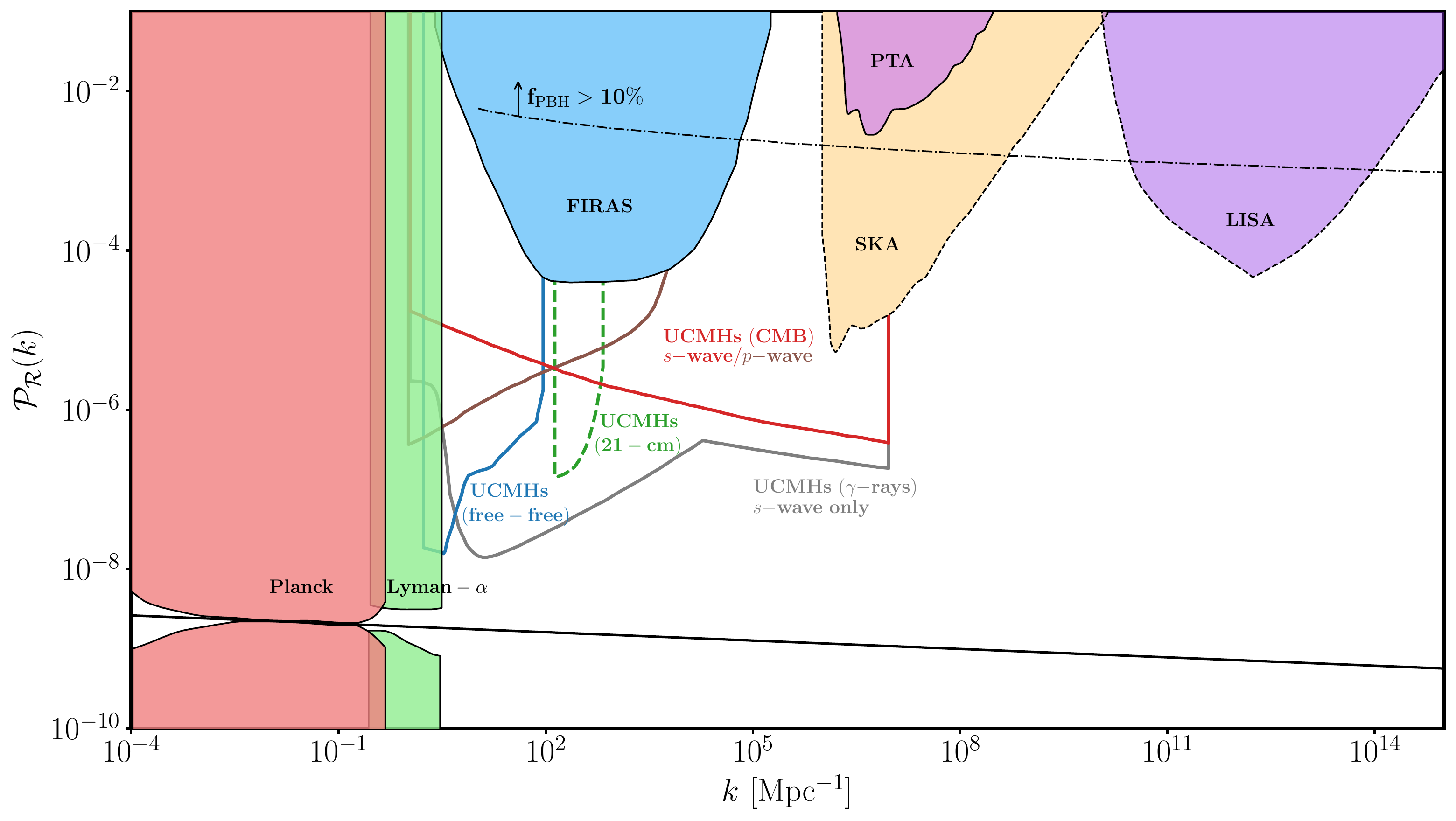} 
\caption{\small Compilation of the most up-to-date constraints on the primordial power spectrum $\mathcal{P}_{\mathcal{R}}(k)$ (solid lines) and sensitivity forecasts for the next generation observatories (dashed lines). Filled areas are obtained independently of UCMHs formation from:  CMB anisotropies \cite{Planck:2018vyg}  (red),  Lyman-$\alpha$ observations \cite{Bird:2010mp} (green),  CMB spectral distortions \cite{Chluba:2012we} (blue), Pulsar Timing Array \cite{Byrnes:2018txb} (pink), SKA forecasts (yellow) and LISA forecasts (purple) \cite{Inomata:2018epa}. Non-filled areas are the consequence of UCMHs forming from a spiky (i.e., monochromatic) power-spectrum enhancement. Our main results from exotic energy injection in the CMB are indicated by the red and brown curves (see \refsec{sec:results} for details). The rest of signals/signatures/observations used here are: an optimistic free-free emission scenario \cite{Abe:2021mcv}  (blue), the 21 cm fluctuations \cite{Furugori:2020jqn} (green), the point-like source search in $\gamma$ rays and the diffuse $\gamma$-ray search \cite{Delos:2018ueo} (gray). The latter and our constraints in red assume $s$-wave annihilating DM (into $b\bar{b}$) with mass $m_\chi = 1 \ \rm{TeV}$ and cross-section $\sigma_0 = 3\times 10^{-26} ~ \rm cm^3 s^{-1}$. Our constraints in brown assume instead $p$-wave DM annihilations with cross-section $\sigma_1 = 10^{-19} ~ \rm cm^3 s^{-1}$. The black dash-dotted line indicates the values of the monochromatic power-spectrum enhancement for which the fraction of DM in the form of PBH becomes larger than $10\%$ (see \refapp{app:prior_region}). The black solid line shows the \Planck~best fit primordial spectrum $\mathcal{P}_{\mathcal{R}}(k)$  of the $\Lambda$CDM model. }
\label{fig:compilation_constraints}
\end{figure}

In this analysis we focus on another consequence that an enhancement of the primordial power spectrum, even if too small to trigger PBH formation, would lead to: the formation of small DM clumps, called ultracompact minihalos (UCMHs), at possibly large redshifts ($z \gtrsim 100$) \cite{Ricotti:2009bs}. The presence of UCMHs can be probed by various methods and therefore indirectly constrain the primordial power spectrum. The heated ionized gas embedded in massive UCMHs would emit radiation via the free-free process imprinting in the diffuse free-free background -- probed by the foreground analysis of the CMB \cite{Abe:2021mcv}. UCMHs could similarly impact on the 21cm temperature fluctuations by clumping baryons at large redshifts \cite{Furugori:2020jqn}. If present in the MW today (because cuspier than traditional halos formed at $z \lesssim 20$  \cite{Delos:2018ueo}), they could also produce detectable astrometric microlensing events \cite{Li:2012qha} or be seen passing through the line of sight to pulsars thanks to pulsar timing monitoring \cite{Clark:2015sha}. If we further assume that DM is made of weakly interacting massive particles (WIMPs) that could annihilate into standard model species, one can further efficiently trace the presence of UCMHs with gamma-ray and neutrino observations \cite{Yang:2013dsa, Nakama:2017qac, Bringmann:2011ut, Josan:2010vn, Scott:2009tu, Delos:2018ueo,Zhang:2021mth}. The UCMHs would act as super-booster of the annihilation rate by more efficiently clumping DM than standard halos. In addition, the extra energy released and deposited in the baryon gas due to the annihilation could also indirectly modify the 21cm signal by changing the thermal history of the Universe \cite{Yang:2016cxm}.  \\

This paper is dedicated to the study of the imprint left on CMB anisotropies by UCMH-induced DM annihilation, strongly boosted with respect to standard CDM halos. We anticipate that UCMHs will have a significant impact due to their steeper profile and earlier formation, potentially allowing them to roam the Universe during the period of maximum sensitivity of the CMB to energy deposition \cite{Zhang:2010cj,Yang:2011ef,Yang:2011jb,Natarajan:2015cva}. 
Our analysis considers both velocity-independent ($s$-wave) and velocity-dependent ($p$-wave) cross-sections. While the latter case is usually not considered with cosmological probes due to the small velocity of the smooth DM component \cite{Diamanti:2013bia}, it can still be relevant here as DM particles in virialised and early collapsed structures have a higher velocity dispersion than that in the homogeneous background. Specifically, we investigate the constraints that can be set on a monochromatic enhancement of the primordial power spectrum,
\begin{equation}
\mathcal{P}_{\rm \star}(k) = \As \ks \delta(k-\ks) \, .
\label{eq:SpikyPower}
\end{equation}
The authors of Ref.~\cite{Kawasaki:2021yek} have obtained similar CMB constraints for $s$-wave annihilating DM particles, using the Boltzmann solver {\tt CAMB} \cite{Lewis:1999bs} -- coupled to the thermal history solver {\tt RECFAST} \cite{Seager:1999km, Seager:1999bc} -- for a power spectrum in the shape of \refequ{eq:SpikyPower}. They have derived the distribution of UCMHs from this monochromatic power spectrum using either the BBKS formalism (also referred to as the peak theory) \cite{Bardeen:1985tr} or the Press-Schechter theory \cite{Press:1973iz}. We improve upon this study in a number of ways. Here we investigate both $s$- and $p$- wave annihilation relying on the Boltzmann solver {\tt CLASS} \cite{Lesgourgues:2011re, Blas:2011rf} and more particularly on the {\tt ExoCLASS} extension including the {\tt DarkAges} module \cite{Stocker:2018avm}. Because {\tt RECFAST} was primarily built with fudge factors to match $\Lambda$CDM model (and thus may have not encompassed the exotic physics scenarios we consider here) \cite{Poulin:2015pna}, within {\tt CLASS} the ionization and thermal history are solved by the more flexible and model independent {\tt HYREC} \cite{Ali-Haimoud:2010hou, Lee:2020obi} module. In addition, we evaluate the impact of late-time structure formation history by considering the sum of the $\Lambda$CDM and spiky power spectra, 
\begin{equation}
\mathcal{P}_{\mathcal{R}}(k) = \mathcal{P}_{\rm PL}(k) + \mathcal{P}_\star(k). 
\end{equation}
We use the excursion set theory (EST) \cite{Bond:1990iw} to predict the distribution of UCMHs, assuming that they gradually disappear from the halo population through accretion and mergers. In practice, this means that we will consider a mixed population of halos, consisting of UCMHs with cuspy density profiles and standard halos with shallower profiles. Although recent studies suggest that UCMHs cusps can survive until present times \cite{Delos:2022yhn}, major mergers \cite{Ishiyama:2014uoa, Ogiya:2016hyo, Angulo:2016qof} and rapid accretion could still impact their size and shape. The presence of ultracompact subhalos would modify the DM clumpiness and velocity distribution which would undoubtedly have consequences on the annihilation boost factor. However, an accurate analytical treatment of this effect within the EST framework would require a recipe to evaluate the ultracompact subhalo mass function in every host as well as their tidal stripping, with all the associated astrophysical complexity. This is beyond the scope of this analysis. Therefore we bracket the uncertainties related to accretion and mergers by also deriving the boost factor with the BBKS formalism \cite{Bardeen:1985tr} following Ref.~\cite{Kawasaki:2021yek}. In this latter approach all density peaks associated with the additional power-spectrum spike are assumed to collapse into UCMHs contributing to the DM annihilation boost factor up to today. This amounts to consider a negligible merger rate, thus leading to more optimistic constraints. We highlight that the theoretical uncertainties associated to the choice of formalism to describe the UCMH population are limited in the redshift range where the CMB is most sensitive but that they can have serious consequences for late time probes (\emph{e.g.,} for the 21cm signal). Additionally, we demonstrate that even the $p$-wave process has constraining power on the power spectrum for cross-sections not yet ruled out by the strongest astrophysical constraints. \\

In \reffig{fig:compilation_constraints} we show our main results and compare them to the strongest constraints in the literature. The upper dash-dotted black line shows the amplitude threshold necessary to reach a DM fraction of 10\% in the form of PBHs. It lies orders of magnitude above the UCMHs-induced constraints. In the $s$-wave scenario, for a thermal dark matter of 1 TeV, the CMB anisotropies set constraints that are comparable to those of astrophysical probes. In the $p$-wave scenario, for a cross-section not yet ruled out by current astrophysical constraints, they set a competitive bound on modes between 10 and 100 $\rm Mpc^{-1}$.  \\

The paper is organized as follows. In \refsec{sec:EnergyInjection}, we discuss the relationship between dark matter annihilation and energy deposition in the intergalactic medium (IGM), along with the quantitative impact on CMB anisotropies. Next, in \refsec{sec:BBKS}, we review the computation of the UCMH-induced annihilation boost using the BBKS formalism. Subsequently, we present our new boost computation based on the EST approach in \refsec{sec:EST_HMF}.
In \refsec{sec:results} we show the results of our numerous analyses and compare to previous work. We conclude in \refsec{sec:conclusion}. Complementary information can be found in various appendices. In \refapp{app:injvel} we provide a careful derivation of the injected energy and boost factor for velocity dependent cross-sections. In \refapp{app:HeavisideInPS} we detail some important aspects about the mass function and formation redshift in the EST approach. In \refapp{app:prior_region} we explain the physical motivation behind our chosen prior for the primordial spike amplitude. Finally, in \refapp{app:k_fs} we evaluate the impact of a finite DM free-streaming scale on the constraints.

\section{Energy deposition and impact on the CMB anisotropies}
\label{sec:EnergyInjection}
In this section, we first calculate the energy injected into the IGM from DM annihilation via $s$- or $p$-wave processes. We then derive the corresponding expression for the deposited energy and describe its implementation in the {\tt DarkAges} package, followed by a discussion of the implications for the cosmic microwave background (CMB) anisotropies.  We relate the distribution of UCMHs (addressed in \refsec{sec:BBKS} and \refsec{sec:EST_HMF}) to the clumping boost function. For the sake of convenience, we adopt the natural units convention ($c = \hbar = k_{\rm B} = 1$) throughout the rest of the paper.

\subsection{Energy injection from velocity-(in)dependent DM annihilation}

We review here the expressions related to exotic energy injection in the IGM due to DM annihilation into standard model (SM) species. Let us denote by  $\sigma(v_{\rm rel})$ the corresponding total cross-section that can depend on the relative velocity between the two incoming particle. The energy injected from that process is proportional to the cross-section modulated by the DM velocity distribution. The latter is characterised by the one point phase-space distribution function (PSDF), $f(\bf v, \bf r)$. The energy injection rate (per units of volume) is therefore given by, 
\begin{equation}
       \left.  \frac{{\rm d} E}{{\rm d} t {\rm d} V} \right|_{\rm inj.} ({\bf r})  =  \frac{1}{m_\chi} \int {\rm d}^3 {\bf v}_1  {\rm d}^3 {\bf v}_2 f({\bf v}_1, {\bf r}) f({\bf v}_2, {\bf r}) |{\bf v}_2 - {\bf v_1}|  \sigma\left(  |{\bf v}_2 - {\bf v_1}| \right)\, .
\end{equation}
In this work, we simplify the analysis of velocity dependencies by only considering the first terms of the partial wave expansion. By doing so, we can express the total cross-section in a simple parametric form given by 
\begin{equation}
\sigma(v_{\rm rel}) v_{\rm rel} = \sum_{\ell =0}^{\infty} \sigma_\ell v_{\rm rel}^{2\ell} \, ,
\end{equation}
where the index $\ell$ refers to the angular momentum of the incoming annihilating particles.  More particularly, we focus our analysis on the $\ell=0$ and $\ell=1$ modes of the annihilation cross-section, which are known as the $s$- and $p$-wave cross-sections, respectively. In realistic particle models, the total annihilation cross-section typically consists of an infinite series of such terms. However, due to the non-relativistic nature of dark matter, $v_{\rm rel} \ll 1$ and the lowest non-zero order term dominates (provided that the factors $\sigma_\ell$ do not explode). As a result, the $s$-wave case has received significant attention in the literature, with a wide range of cosmological and astrophysical probes used to constrain it. The current constraints exclude a thermally produced dark matter particle with a mass below $\sim$100 GeV \cite{Planck:2018vyg, Fermi-LAT:2015att}. However, some particle models naturally predict a vanishing $s$-wave term and a dominant $p$-wave term. These scenarios are much less constrained by indirect searches due to the large suppression by the velocity factor. An example is provided by fermionic dark matter particles coupled to the Standard Model species through a scalar mediator, as in Higgs-scalar portal models. For reviews on simplified models featuring $p$-wave annihilation cross-sections and their connections to UV complete models, see Refs. \cite{Arina:2018zcq, DeSimone:2016fbz,Abdallah:2015ter}. 

In \refapp{app:injvel_dEdt} we derive the general expression (for any value of $\ell$) of the energy density injection rate associated to an isotropic velocity distribution of particles, that is, assuming $f({\bf v}, {\bf r}) = f(v, {\bf r}) $ with $v \equiv|{\bf v}|$. We show that it is a finite sum over products of the moments of the velocity norm. By convention, the PSDF is normalised to the DM density, $\rho_\chi(\bf r)$, when integrated over the entire range of velocity. These moments are thus defined by
\begin{equation}
    \left< v^n \right>({\bf r}) \equiv \frac{1}{\rho_\chi({\bf r})}\int f(v, {\bf r}) v^{n} {\rm d}^3 {\bf v}\, , \quad \forall n\in \mathbb{N}\, .
    \label{eq:velocity_moment}
\end{equation}
In the $s$ and $p$-wave scenarios the expressions of the energy density injection rate reduce to the following formula,
\begin{equation}
    \left.  \frac{{\rm d} E_\ell}{{\rm d} t {\rm d} V} \right|_{\rm inj.} ({\bf r})  =  \left[2\left< v^2\right>\right]^\ell \rho_\chi^2({\bf r}) \frac{\sigma_\ell}{m_\chi} \, .
    \label{eq:dEl_dtdV}
\end{equation}

We define $\mathcal{R}_{\ell}(z)$ as the ratio of the actual spatially-averaged injection rate to the injection rate assuming a homogeneous, i.e., smooth universe, and an annihilation cross-section that is purely $s$-wave. Let us denote the present-day average density of dark matter by $\overline{\rho_{\chi,0}}$, and we use this notation to set
\begin{equation}
    \begin{split}
    \left.  \overline{\frac{{\rm d} E_\ell}{{\rm d} t {\rm d} V} } \right|_{\rm inj.} & \equiv   \mathcal{R}_\ell(z)  \frac{\sigma_\ell}{\sigma_0}  \left.  \frac{{\rm d} E_0}{{\rm d} t {\rm d} V} \right|_{ \rm  inj., sm} =  \mathcal{R}_\ell(z) \frac{ \sigma_\ell}{m_\chi} \overline{\rho_{\chi, 0}}^2(1+z)^6 \, .
    \label{eq:Def_Rl}
    \end{split}
\end{equation}
Hence, the velocity dependence of the cross-sections and the impact of DM clumping into halos are effectively encapsulated in $\mathcal{R}_\ell(z)$. On the contrary, the ratio of $\sigma_\ell/ \sigma_0$ introduced in the definition ensures that it is independent of the particle model. In the subsequent subsection, we provide a more comprehensive expression for this quantity.

\subsection{Boost factor from DM clumping into halos}

In the halo model, the total density of DM at each point in the Universe can be written as the sum of a smooth background and a clumpy distribution of halos. Therefore, $\mathcal{R}_\ell(z)$ can be expressed in terms of what we call an effective halo boost function, denoted as $\mathcal{B}_\ell(z)$\footnote{It's worth noting that we refer to this boost as "effective" since, in the case of $p$-wave annihilation, it can take values lower than 1 - this is due to the fact that $\mathcal{R}_\ell(z)$ is defined with respect to the $s$-wave annihilation rate.}. In \refapp{app:injvel_sp}, starting from \refequ{eq:dEl_dtdV} and \refequ{eq:Def_Rl} we obtain that for $\ell=0$ or $\ell =1$, 
\begin{equation}
\begin{cases}
    \displaystyle \mathcal{R}_0(z) = 1 - \left(1-\frac{\rho_{\rm sm}(z)}{\overline{\rho_{\chi}}(z)}\right)^2 + \mathcal{B}_{0}(z)\\[10pt]
    \displaystyle \mathcal{R}_1(z) \simeq 2\left(\frac{\rho_{\rm sm}(z)}{\overline{\rho_\chi}(z)}\right)^2\left<v^2\right>_{\rm sm}  +  2\frac{\rho_{\rm sm}(z)}{\overline{\rho_\chi}(z)} \left\{1-\frac{\rho_{\rm sm}(z)}{\overline{\rho_\chi}(z)}\right\}\left[\left<v^2\right>_{\rm sm} + \overline{\left<v^2\right>}_{\rm h}\right] + \mathcal{B}_{1}(z) \, .
    \end{cases}
    \label{eq:Rlsimpl}
\end{equation}
We have introduced here $\rho_{\rm sm}$, the DM density in the smooth background (\emph{i.e.}, the total DM density to which we subtract the average DM density in halos). This expression also involves the velocity dispersion $\left< v^2 \right>_{\rm sm}(z)$ of the smooth DM background and the average velocity dispersion in halos $\overline{\left<v^2\right>}_{\rm h}$ -- see the appendix for more details. Assuming that the DM is thermally coupled to the primordial plasma up to redshift $z_{\rm kd}$ when the plasma reached the kinetic decoupling temperature $T_{\rm kd}$ \cite{Diamanti:2013bia},
\begin{equation}
    \left< v^2 \right>_{\rm sm}(z) = 3\left(\frac{1+z}{1+z_{\rm kd}}\right)^2 \frac{T_{\rm kd}}{m_\chi} = 1.2 \times 10^{-16} \left(\frac{1+z}{1000}\right)^2 \left(\frac{\rm MeV}{T_{\rm kd}} \right) \left( \frac{\rm GeV}{m_\chi} \right)
\end{equation}
In realistic WIMP models, the kinetic decoupling temperature is of the order MeV for GeV particles \cite{Facchinetti:2021wek} and following the scaling $T_{\rm kd}\sim m_\chi^{2/3}$ from Refs.~\cite{Bringmann:2009vf, Chen:2001jz} the velocity dispersion in the smooth background is always orders of magnitude lower than the speed of light, $\left< v^2 \right>_{\rm sm}(z) \ll 1$. In the $p$-wave case, all the terms involving the smooth background velocity dispersion are then negligible. The remaining term depending on the average velocity dispersion in halos is insignificant before halos are formed as well as when they dominate the DM density. In the $s$-wave case, we can check that either $\overline{\rho_\chi}(z) = \rho_{\rm sm}(z)$ when halos have not formed or $\mathcal{B}_0(z) \gg 1$ as soon as they have. Therefore, in both scenarios \refequ{eq:Rlsimpl} can be approximated by the much simpler expressions
\begin{equation}
    \begin{cases}
        \mathcal{R}_0(z) \simeq 1 + \mathcal{B}_0(z) \\
        \mathcal{R}_1(z) \simeq  \mathcal{B}_1(z) \, .
    \end{cases}
    \label{eq:Rlsimplsimpl}
\end{equation}

To compute the effective boost functions $\mathcal{B}_\ell(z)$, we rely on a semi-analytical method based on the halo model (for other approaches based on numerical simulations, see \cite{Takahashi:2021pse,Sefusatti:2014vha}). In this framework, the halo distribution is characterised by the mass-function or as we introduce here, the co-moving number density in units of mass and formation redshift, $\partial^2 n/(\partial M \partial z_{\rm f})$. This quantity can be directly related to the primordial power spectrum as will be shown in \refsec{sec:BBKS} and \refsec{sec:EST_HMF}. The total boost function $\mathcal{B}_\ell(z)$ takes the form
\begin{equation}
\begin{split}
    & \qquad \, \mathcal{B}_{\ell}(z) = \frac{1}{\overline{\rho_{\rm m, 0}}}\int M \frac{\partial^2 n(z)}{\partial M \partial z_{\rm f}} \mathcal{B}_{\ell, \rm h}(z) {\rm d} z_{\rm f} {\rm d} M \, , \\
    & {\rm with} \quad \mathcal{B}_{\ell, \rm h}(z) \equiv \frac{1}{M\overline{\rho}_{\rm m}(z)} \int_{0}^{r_\Delta(z)} {\rm d}^3 {\bf r} \rho_{\rm h}^2(r, z)\left[2 \left< v^2 \right>_{\rm h}(r, z) \right]^{\ell} \, ,
    \label{eq:BoostFactorDefinition}
\end{split}
\end{equation}
the contribution of a single halo of mass $M$ and formed at redshift $z_{\rm f}$. We refer to the latter quantity as the 1-halo boost, normalised to the averaged matter density $\overline{\rho_{\rm m}}(z)= \overline{\rho_{\rm m, 0}}(1+z)^3$. The properties of spherical halos are characterized by their density profile, $\rho_{\rm h}(r, z)$, which depends on the redshift, mass, and formation redshift. The one-halo boost factor is also influenced by the size of the halo, as determined by the upper bounds of the integral, as well as the velocity-dispersion profile. The virial radius, defined as $r_\Delta^3(z) \equiv 3M/(4\pi \Delta \rho_{\rm c}(z))$, relates the halo's mass to a fixed critical overdensity $\Delta$, with $\rho_{\rm c}(z) = 3H^2(z)/(8\pi G_{\rm N})$ denoting the critical density of the Universe with $G_{\rm N}$ being Newton's constant. We use $\Delta = 200$ since it approximates\footnote{In the spherical collapse model $\Delta = 18\pi^2\sim 178$.} the average overdensity of halos after virialization \cite{Bryan:1997dn}. Finally, assuming spherical halos and isotropic velocity distributions, the velocity-dispersion profile $\left< v^2 \right>_{\rm h}(r, z)$ can be obtained via the Jeans equation and written as
\begin{equation}
    \left< v^2 \right>_{\rm h}(r, z) = \frac{12 \pi G_{\rm N}}{\rho_{\rm h}(r, z)}\int_{r}^{r_\Delta(z)}{\rm d} r' \int_0^{r'}  {\rm d} r'' \left( \frac{r''}{r'} \right)^2 \rho_{\rm h}(r', z)  \rho_{\rm h}(r'', z)   \, .
    \label{eq:VelocityDispersionJeans}
\end{equation}
As shown in \refapp{app:injvel}, Eqs.~(\ref{eq:BoostFactorDefinition}) and (\ref{eq:VelocityDispersionJeans}) can be combined and simplified into a  more compact expression for the $p$-wave case (where the velocity dependent term plays a role).

Let us now detail the one-halo boost expression for each halo in the next subsection.

\subsection{One-halo boost from halo properties}
\label{sec:one_halo_boost}

\begin{figure}[t!]
\centering
\includegraphics[width=0.6\linewidth]{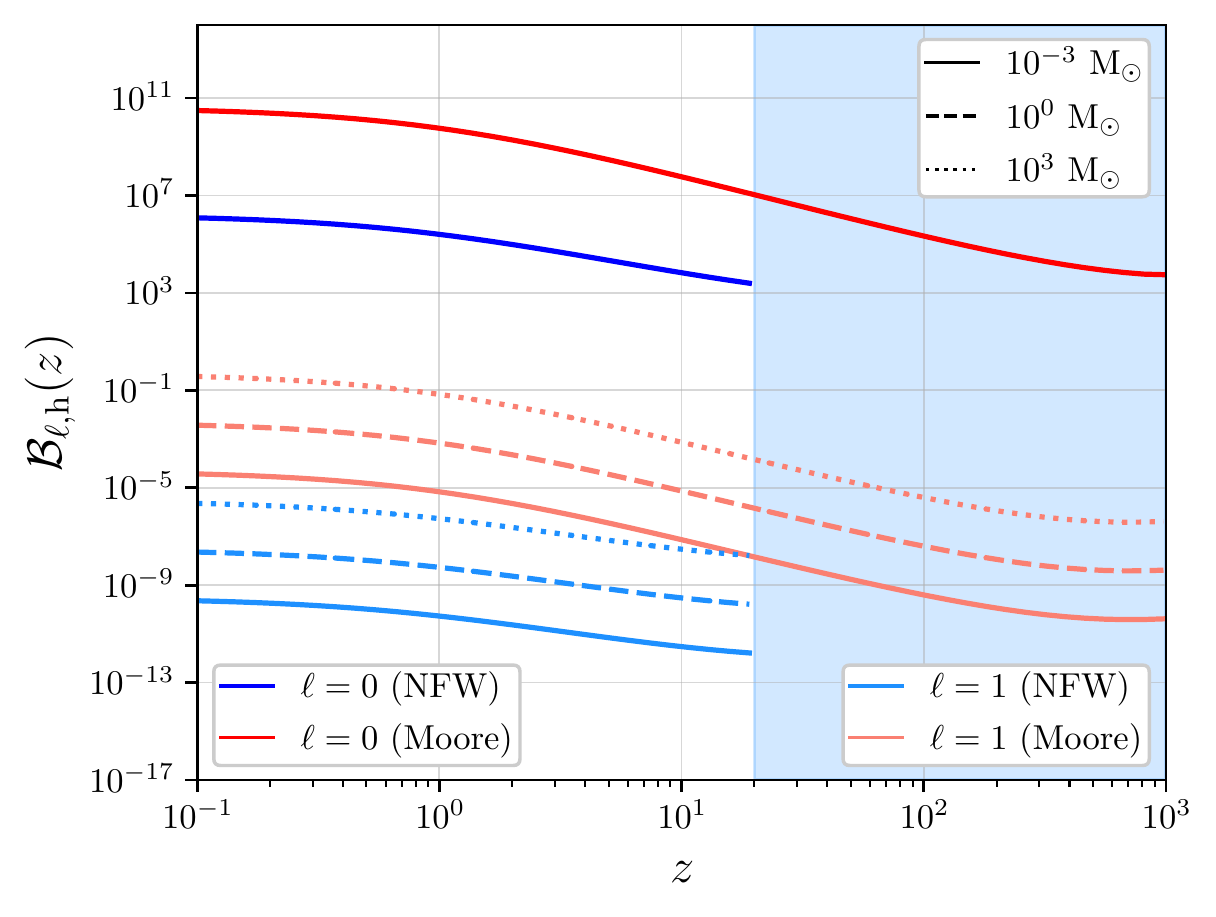} 
\caption{One-halo boost factor in terms of the redshift for three different masses $m_{200}$. The case of an NFW halo is shown in dark blue (for $\ell =0$) and light blue (for $\ell=1$) assuming a concentration $c_{200} = 60$ at redshift 0 (corresponding to a formation redshift $z_{\rm f} = 20$ in the model of Ref.~\cite{Maccio:2008pcd}, see \refsec{sec:EST_HMF_Properties}). The shaded blue area shows indicates when this halo is not yet formed. The case of a Moore halo is shown in dark red (for $\ell=0$) and in light red (for $\ell=1$) considering a scale density $\rho_{\rm s} = 1.2\times 10^{12}~ \rm M_\odot kpc^{-3}$ (corresponding to a formation redshift $z_{\rm f}=1000$ in the model of Ref.~\cite{Delos:2018ueo}, see \refsec{sec:BBKS_HMF_Properties}). The configuration $\ell=0$, Moore profile is regularised assuming $\sigma_0 = 3\times 10^{-26}~\rm cm^3 s^{-1}$ with $m_\chi = 1~\rm TeV$.}
\label{fig:OneHaloBoost}
\end{figure} 

The halo density profile can be written in terms of a dimensionless profile function $\tilde \rho$ and two scalars, the scale radius $r_{\rm s}$ and scale density $\rho_{\rm s}$, such that
\begin{equation}
     \rho_{\rm h}(r, z) = \rho_{\rm s}(z) \tilde \rho \left(\frac{r}{r_{\rm s}(z)}\right) \, .
\end{equation}
In the following, we will work with generalized profiles defined as
\begin{equation}
   \tilde \rho(x) \equiv \frac{1}{x^{\gamma}(1+x)^{3-\gamma}} \quad \forall \gamma \in [0, \infty) \, .
   \label{eq:generlized_NFW}
\end{equation}
In particular we will work with standard NFW \cite{Navarro:1995iw} ($\gamma=1$) and Moore \cite{Moore:1999gc} ($\gamma = 3/2$) profiles. The latter has been shown to adequately describe UCMHs based on numerical simulations \cite{Gosenca:2017ybi, Adamek:2019gns, Delos:2017thv}, as it will be discussed in  \refsec{sec:BBKS_HMF_Properties}. The two scale parameters can also be traded for the equivalent system of \emph{virial} parameters, combination of the concentration $c$ and the virial mass $m$ -- \emph{i.e.,} \emph{cosmological} mass. Virial and scale parameters are related by 
\begin{equation}
\begin{cases}
    \displaystyle c \equiv \frac{1}{r_{\rm s}}\left(\frac{3m}{4\pi \Delta \rho_c}\right)^{1/3} \\
    \displaystyle  m \equiv 4\pi \rho_{\rm s} r_{\rm s}^3\mu(c) 
    \end{cases} \, ,
    ~
    {\rm with}
    \quad \mu(x) \equiv  \int_{0}^{x} \tilde \rho(y) y^2{\rm d}y
    \label{eq:definition_virial_parameters}
\end{equation}
the dimensionless enclosed mass profile. The first equality defines the concentration as a ratio between the virial radius and the scale radius \footnote{Another equivalent convention is to define the concentration, not with respect to the scale radius but to the radius at which the logarithmic slope of the profile is equal to -2. This second option only brings further mathematical complications for our study therefore we choose the definition using the scale radius.}. The second equality relates the virial mass to the halo mass enclosed in the virial radius. From the concentration, one can also define a virial dimensionless luminosity by
\begin{equation}
    \lambda_{\ell}(x) \equiv  \int_0^{x} \tilde \rho^2(y) \left[\frac{1}{\tilde \rho(y)}\int_y^{c} \frac{\tilde \rho(y')\mu(y')}{y'^2} {\rm d} y' \right]^\ell y^2 {\rm d} y \, .
\end{equation}
In the end, using the definition of the virial parameters, the one halo-boost can be written as a product of dimensionless factors
\begin{equation}
    \mathcal{B}_{\ell, \rm h}(z) =  \frac{\Delta}{3\Omega_{\rm m}(z)} \left[2 v^2_\Delta(z)\right]^{\ell
} \frac{c^{3+\ell}\lambda_{\ell}(c)}{\mu^{2+\ell}(c)} \, ,
\label{eq:dimensionlessOneHaloBoost}
\end{equation}
with the virial velocity dispersion
\begin{equation}
\begin{split}
    v^2_\Delta(z) & \equiv 4\pi G_{\rm N} \Delta \rho_{\rm c}(z) r_{\Delta}^2(z) \\
    & \propto (1+z) [\Omega_{\rm m}(z)]^{-1/3} m^{2/3} \, .
    \end{split}
\end{equation}

However, it is easy to show that, in the $s$-wave annihilation configuration, the 1-halo boost computed for a Moore profile is divergent due to the sharp cusp towards the center: $\rho_{\rm h}(r)\propto r^{-3/2}$. In practice, because $s$-wave annihilation is efficient to annihilate particles, we can regularise the divergence by considering the formation of an inner core with a size set from the annihilation efficiency. More precisely the profile is truncated for densities above
\begin{equation}
    \rho_{\rm max} = \frac{m_\chi}{\sigma_0 [t(z) - t(z_{\rm f})]}
    \label{eq:regularisation_density_profile}
\end{equation}
where $t(z)$ in the cosmic time at redshift $z$. For higher order annihilation processes, such as $p$-wave, the 1-boost factor is non divergent as the integral is regularised by the velocity dispersion (low in the center). In turn, the annihilation in the inner part of the halos is then not efficient enough to result in a core formation so that no truncation is needed.\\

In \reffig{fig:OneHaloBoost}, we show the one-halo boost in terms of the redshift and for three fixed masses. Two scenarios are considered, an NFW halo formed at redshift $z_{\rm f} = 20$ and a Moore halo formed at redshift $z_{\rm f} = 1000$. Formation redshift and halo parameters are related using fits on dedicated numerical simulations and discussed in \refsec{sec:BBKS_HMF_Properties} and \refsec{sec:EST_HMF_Properties}. Early formed halos with a Moore profile (what corresponds to UCMHs) shows a higher boost than \emph{standard} NFW halos formed at much later time. Here, the Moore profile is regularized assuming $\sigma_0 = 3\times 10^{-26}~\rm cm^3 s^{-1}$ with $m_\chi = 1~\rm TeV$, a different choice would re-scale the red curve -- see \refequ{eq:l0_moore}. The boost primarily increases at low redshifts due to the decreasing background matter density, with a rough scaling of $\mathcal{B}_{\ell,\rm h}\propto (1+z)^{-3+\ell}$. In the $s$-wave case, the one-halo boost is mass-independent, while the $p$-wave boost scales as $m^{2/3}$. This scaling limits the impact of $p$-wave annihilation coupled to large mode enhancements of the power spectrum, which only produce UCMHs with mass $ m \sim \ks^{-3}$ and boosts $\mathcal{B}_{1, \rm h} \sim k_\star^{-2}$. Consequently, the constraints on $\As$ show a steep dependence on $k_{\star}$ in the $p$-wave scenario. \\

We have now detailed all the ingredients to evaluate the injection rate of energy density in the presence of UCMHs. To conclude this section on the theoretical framework we address, in the following subsection, the relation between the energy injected and the energy deposited in the baryon gaz as well as the consequences for the CMB anisotropies.

\subsection{From energy injection to energy deposition}

The injected energy is not directly deposited into the IGM but through the interactions of the electromagnetic cascade with the baryons in the gaz (in particular, with atoms of hydrogen and sub-dominantly of helium). These interactions modify the evolution of the free electron fraction $x_e \equiv n_e/n_H$ as well as the IGM temperature $T_M$ (which has a feedback on the evolution of $x_e$). Three deposition channels are possible: ionization, excitation of the Lyman-$\alpha$ transition, and heating of the IGM
\footnote{In reality, there is also a fourth ``lost'' photon channel, arising when the energy of extra particles drops below the Lyman-$\alpha$ transition energy ($10.2 \ \rm{eV}$), and are no longer able to interact with atoms in the plasma.}. All of them are not equally efficient according to the energy spectrum of the injected particles. The baryon gas is transparent in some energy ranges. This can be accounted for through deposition functions $f_{\ell, c}(z)$ (where $c$ = heat, ion., exc. refers to the deposition channels) such that
\begin{equation}
     \left.  \frac{{\rm d} E_{\ell}}{{\rm d} t {\rm d} V} \right|_{\rm c, dep.} = \frac{\sigma_\ell}{\sigma_0} f_{\ell, c}(z)   \left.  \frac{{\rm d} E_0}{{\rm d} t {\rm d} V} \right|_{\rm inj., sm} 
\end{equation}
and computed by \cite{Poulin:2015pna, Diamanti:2013bia}
\begin{equation}
\begin{split}
    & f_{\ell, c}(z) \equiv \frac{1}{N_{\ell, c}(z)}  \int \frac{(1+z')^2 {\rm d} z'}{H(z')} \mathcal{R}_\ell(z') \sum_{p} \int T_{c, p}(z', z, E) E \left. \frac{{\rm d}N_p}{{\rm d} E} \right|_{\rm inj.} {\rm d} E \, \\
    & \qquad \text{with} \quad     N_{\ell, c}(z) \equiv \frac{(1+z)^3}{H(z)}\mathcal{R}_\ell(z)\sum_p \int E \left. \frac{{\rm d}N_p}{{\rm d} E} \right|_{\rm inj} {\rm d} E \, ,
\end{split}
\label{eq:f_of_z}
\end{equation}
the normalisation factor. The sum over $p$ represents the sum over the different primary particles produced in the annihilation (typically $p=e^{\pm},\gamma$) and ${\rm d} N_p/ {\rm d} E |_{\rm inj}$ is the injected spectrum of each primary particles. The transfer functions $T_{c, p} (z',z,E)$ give the fraction of $p$-particle's energy $E$ injected at $z'$ (through the channel $c$) that is absorbed at $z < z'$. The calculation of these transfer functions is very involved, but it has been carefully done in \cite{Slatyer:2009yq,Galli:2013dna,Slatyer:2015kla}. To simplify the calculation, the \textsl{on-the-spot} approximation has usually been adopted in the past, which amounts to assuming that
the energy injected at a certain redshift is absorbed at the same redshift, $z' \simeq z$. In this picture, the function $f_c(z)$ is factorized into a ``universal'' deposited energy function $f(z)$ (often taken to be constant, $f_{\rm eff}\equiv f(z_{\rm eff})$) and coefficients $\chi_c (x_e)$ describing the repartition in each channel \cite{Chen:2003gz,Galli:2013dna}. While this approximation has been shown to work well for DM annihilations in the smooth background, it fails to capture the redshift-dependence induced by the formation of DM halos \cite{Poulin:2015pna}.\ 

For this reason, we perform a full modelling of the energy deposition by DM annihilations in UCMHs using our modified version of {\tt ExoCLASS} \footnote{Our modified {\tt ExoCLASS} version is publicly available at \url{https://github.com/GuillermoFrancoAbellan/ExoCLASS}. } \cite{Stocker:2018avm}. This code calls the {\tt DarkAges} module, whose main role is to compute the integral in \refequ{eq:f_of_z} using the most up-to-date transfer functions from \cite{Slatyer:2015jla,Slatyer:2015kla} 
\footnote{One of the main assumptions in the computation of these transfer functions is that the modified free electron fraction will not significantly back-react on the energy cascade evolution, which is expected to be a good approximation given the tight CMB constraints on any kind of exotic energy injection.}. Once the quantities ${\rm d} E_{\ell}/({\rm d} t {\rm d} V)|_{\rm c, dep.}$ have been obtained, the code solves for the modified recombination history by calling to either {\tt RECFAST} \cite{Seager:1999km} or {\tt HYREC} \cite{Lee:2020obi}. We note that {\tt RECFAST} relies on some fitting functions in order to account for the effect of highly-excited states, which would have to be extrapolated beyond their range of validity in the case of DM halos. For this reason, we choose to use {\tt HYREC}, where those effects are computed at a more fundamental level without necessitating the interpolation \cite{Poulin:2015pna}. While correctly describing
those highly-excited state corrections,  {\tt HYREC} (like {\tt RECFAST}) still considers a system of coupled differential equations having the same structure as Peebles \textsl{three-level atom} model. Schematically, the code integrates
\begin{align}
\frac{dx_e}{dz} &= \frac{1}{(1+z)H(z)} (R(z) - I(z) - I_{X_i}(z)- I_{X_\alpha}(z)),   \\
\frac{dT_{\rm M}}{dz} &= \frac{1}{(1+z)H(z)} \left[   2T_{\rm M} + \gamma (T_{\rm M}-T_{\rm CMB}) \right] + K_h(z).
\end{align}
Here $R$ and $I$ denote the standard recombination and ionization rates, while $I_{X_i}$, $I_{X_\alpha}$ and $K_h$ denote the ionization, excitation and heating rates coming from DM annihilations, which are directly proportional to ${\rm d} E_{\ell}/({\rm d} t {\rm d} V)|_{\rm c, dep.}$ (we refer the reader to  \cite{Poulin:2015pna} for the explicit expressions of those rates). \

\begin{figure}[t!]
\centering
\includegraphics[width=0.49\linewidth]{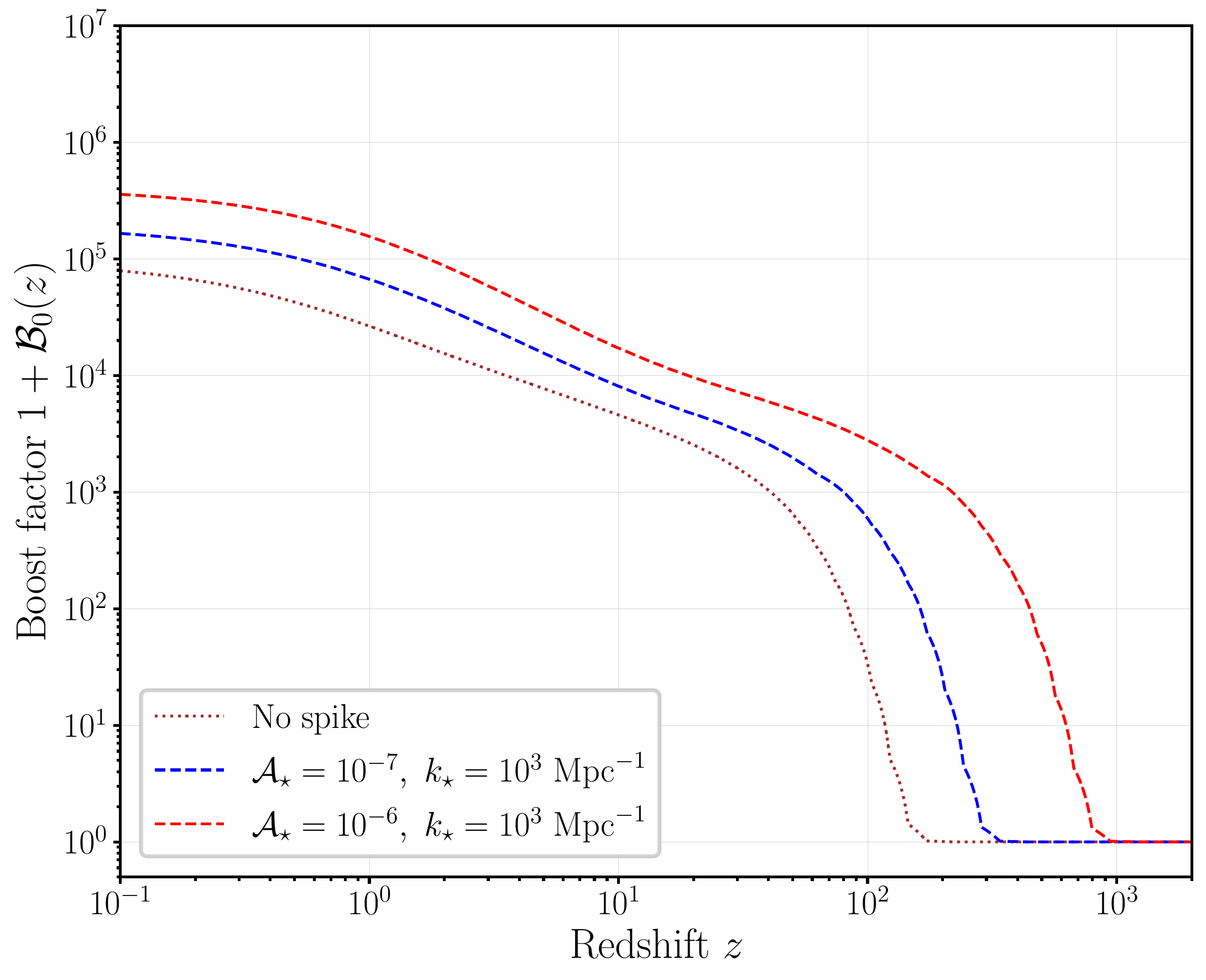} \hfill
\includegraphics[width=0.49\linewidth]{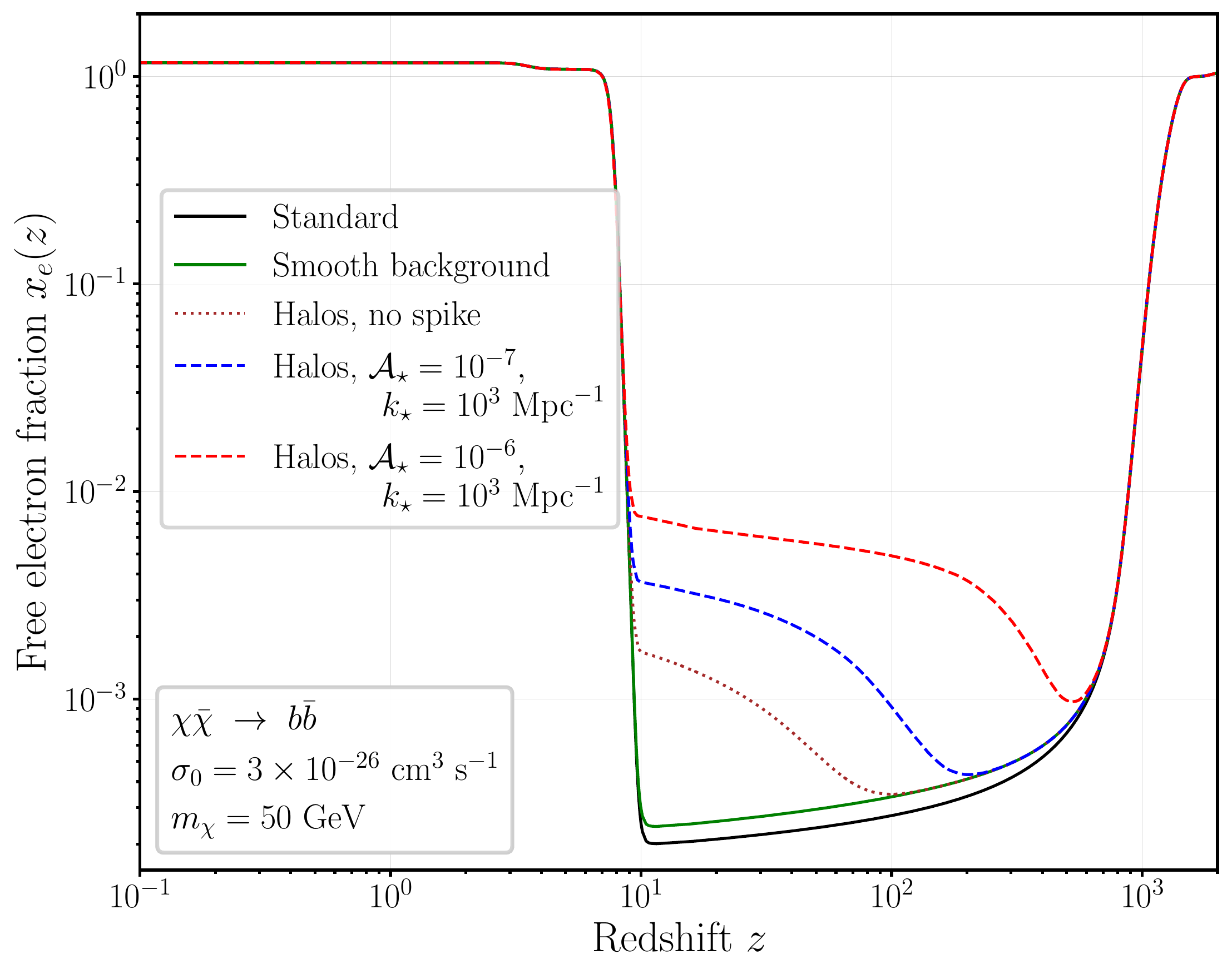}
\caption{ {\bf Left panel.} Evolution of the boost factor for $s$-wave DM annihilations, computed using EST and considering both the smooth $\Lambda$CDM and the spiky components in the primordial power spectrum (see \refsec{sec:EST_HMF}). The dotted line corresponds to the case without the spiky component (i.e., only standard halos), whereas the blue and the red dashed lines consider a spike with $\mathcal{A}_{\star} =10^{-7}$ and $\mathcal{A}_{\star} =10^{-6}$, respectively (both at $k_{\star} = 10^3 \ \rm{Mpc}^{-1}$.) {\bf Right panel.} Evolution of the free electron fraction for the same halo models as in the left panel, under the assumption that DM annihilates into $b\bar{b}$ with a mass $m_{\chi} = 50 \ \rm{GeV}$ and a thermal relic cross-section. For comparison, we also show the case of DM annihilations into the smooth background (green solid line) and no DM annihilations (black solid line). The rest of cosmological parameters are fixed to the \Planck~best fit of $\Lambda$CDM model. }
\label{fig:boost_and_xe}
\end{figure}

\begin{figure}[t!]
\centering
\includegraphics[width=0.49\linewidth]{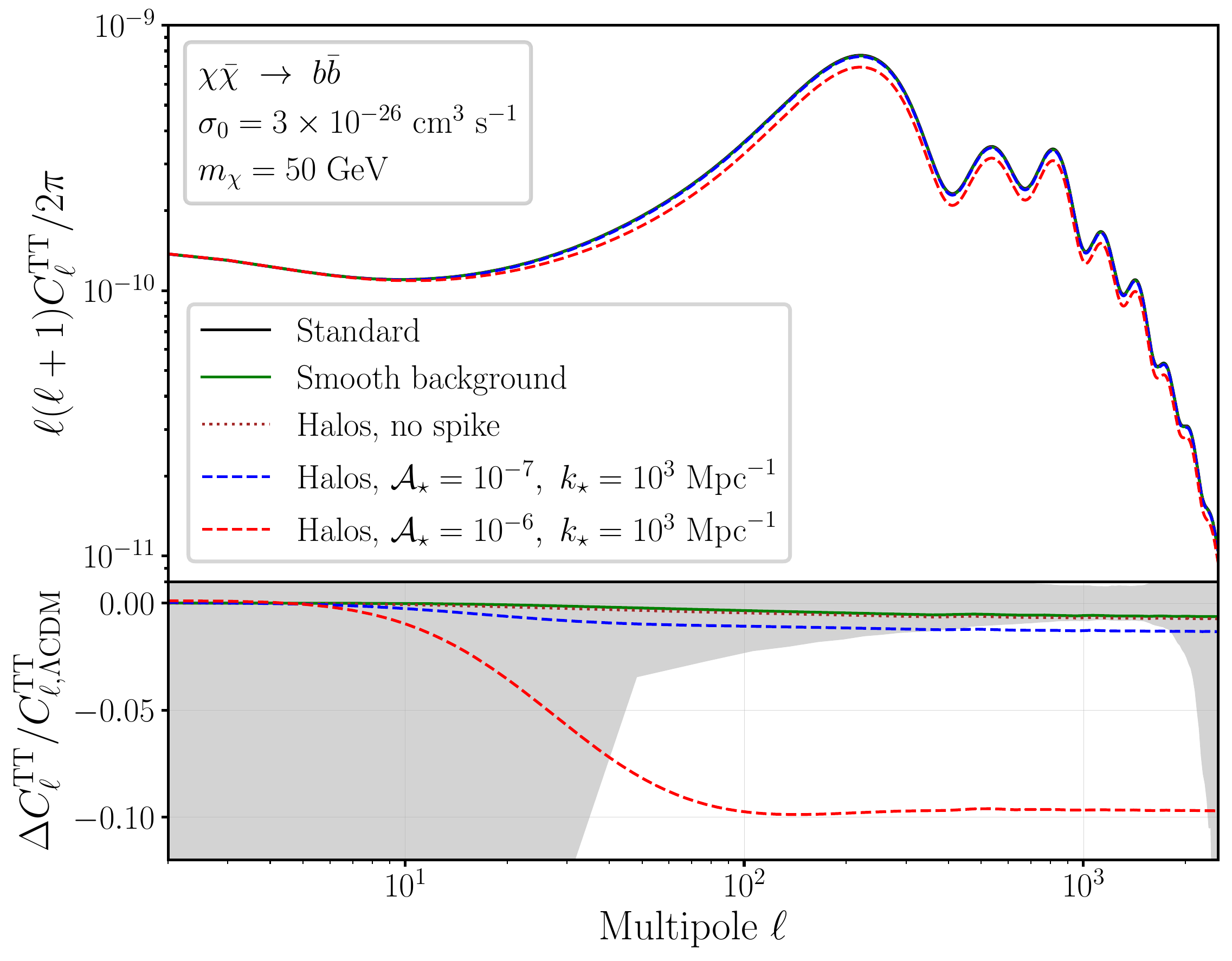} \hfill
\includegraphics[width=0.49\linewidth]{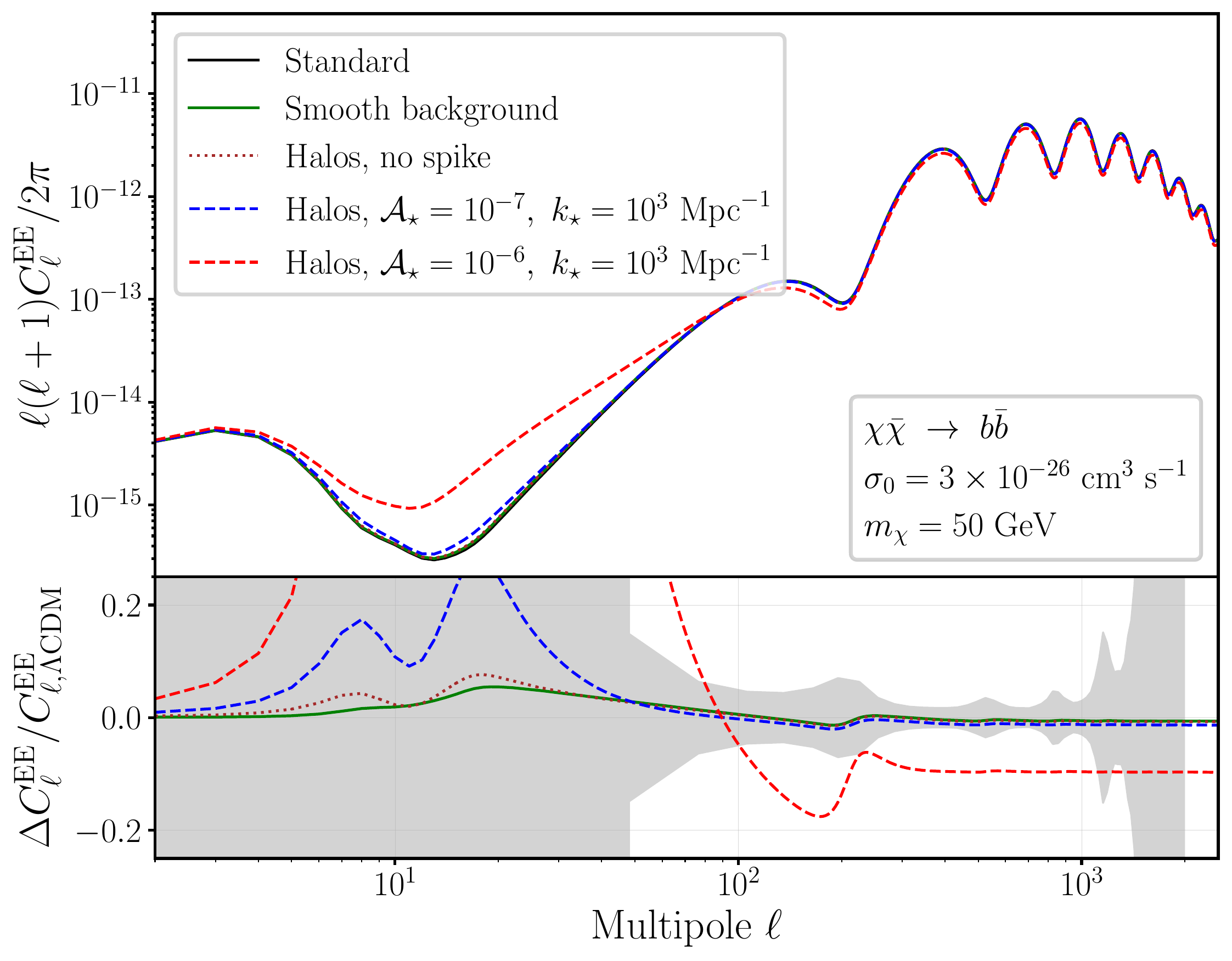}
\caption{ {\bf Left panel.} CMB TT power spectrum for the same models considered in \reffig{fig:boost_and_xe}, together with the residual differences computed with respect to $\Lambda$CDM.  {\bf Right panel.} Same as in the left pannel, but for the CMB EE power spectrum. The gray shaded regions indicate \Planck~$1\sigma$ uncertainties. The rest of cosmological parameters are fixed to the \Planck~best fit of $\Lambda$CDM model.  }
\label{fig:TT_and_EE}
\end{figure}

In general, DM annihilation leave an impact on the CMB anisotropy spectra through a delayed recombination, an increase in the width of the last scattering surface, and an enhancement in the number of residual free electrons after decoupling\footnote{In mathematical terms, the modified free electron fraction $x_e$ affects the Thomson optical depth $\tau(z)$ and the visibility function $g(z)$, two important quantities entering in the line-of-sight solution of the CMB photon Boltzmann hierarchy \cite{Seljak:1996is}.}. The increased number of residual free electrons results in more Thomson scatterings and leads to two main effects: the damping of anisotropies at $\ell > 10$, and the generation of anisotropies around $\ell \sim 30$ in polarization. These effects are small if $x_e$ remains close to its asymptotic freeze-out value, but can be important when $x_e$ starts raising due to the formation of stars or DM halos.  In the presence of UCMHs, the DM annihilation signal is boosted at very high $z$ (well before star reionization), leading to larger $x_e$ and even stronger effects on the CMB temperature and polarization spectra. This can in turn be used to place constraints on $\As$ and $\ks$. In \refsec{sec:BBKS} and \refsec{sec:EST_HMF} we will discuss how to compute the boost function following two different approaches: the BBKS formalism (considering only the spiky component in $\mathcal{P}_{\mathcal{R}}$) and the EST formalism (considering both the smooth $\Lambda$CDM and spiky component in $\mathcal{P}_{\mathcal{R}}$). Before entering into the details of these calculations, it is interesting to illustrate how UCMHs affect the thermal history of the universe and the CMB spectra.  \ 

In the left panel of \reffig{fig:boost_and_xe} we show the boost factor for $s$-wave DM annihilations computed with the EST formalism, for different values of $\As$. We see that for $\As = 0$ (i.e., no spike), the boost factor starts increasing around $z \sim 100$, which is the approximate redshift at which standard halos begin to form. Once the spike is added, the boost starts increasing at much higher $z$ due to early UCMH formation, and it can even become significant close to recombination $z \sim 1000$ for sufficiently high values of $\As$. On the right panel of \reffig{fig:boost_and_xe} we show the corresponding effect on the free electron fraction, for a model in which DM annihilates into $b\bar{b}$ with a mass $m_{\chi} = 50 \ \rm{GeV}$ and a thermal relic cross-section $\sigma_0  =  3 \times 10^{-26} \ \rm{cm}^3 s^{-1}$ (which is currently allowed by standard \Planck~constraints \cite{Planck:2018vyg}). For comparison, we also show the case of DM annihilations in the smooth background and the case of no DM annihilations. As expected, the residual fraction of free electrons after decoupling starts increasing significantly in the presence of halos, with the intensity of this effect being controlled by $\As$. In \reffig{fig:TT_and_EE} we show the impact on the temperature and polarization CMB spectra for the same models. We observe the various effects that we mentioned previously (like the damping of anisotropies at high-$\ell$), with varying degrees of intensity. In particular, the case of standard halos leads to a very tiny effect, comparable to that of annihilations in the smooth background. On the contrary, the impact of early-forming UCMHs is so strong that it is already ruled out by \Planck~error bars. This clearly illustrates the interesting potential of the CMB for constraining the presence of UCMHs.\ 

Let us end this section by adding that, in principle, one could also use CMB data to constrain the impact of DM annihilations within UCMHs during the epoch of reionization. This could be done by combining the predicted DM annihilation signal in UCMHs with some (physically well-motivated) parametrization of the reionization from stars. In practice, the astrophysical source terms suffer from large theoretical uncertainties (but see \cite{Abe:2022swi}). In addition, previous studies have shown that the CMB spectra are rather unsensitive to the entire reionization history $x_e(z)$, since the high-$\ell$ part mainly probes the integrated parameter $\tau_{\rm reio}$, i.e., the optical depth at reionization \cite{Poulin:2015pna}. For these reasons, we decide to follow the same prescription as the \Planck~collaboration, and simply paste the modified evolution of $x_e(z)$ at high-$z$ with the usual ``single-step reionization'' \cite{Lewis:2008wr} at low-$z$\footnote{This parametrization for $x_e(z)$ considers a half-hyperbolic tangent centered on $z_{\rm reio}$ with a very narrow width $\Delta z = 0.5$, that asymptotically reaches $x_e(z=0) = 1$ (if He is neglected).}. We anticipate that future tomographic surveys of the 21cm observable (sensitive to both $x_e(z)$ and $T_{\rm M}(z)$) will shed more light on the interplay between DM annihilations in UCMHs and the stellar sources of reionization. In particular, these surveys will be sensitive to the effects of late-time halo mergers, for which we believe that our improved EST formalism (presented in \refsec{sec:EST_HMF}) should be used.

\section{Properties of UCMHs in the BBKS approach}
\label{sec:BBKS}

In this section we relate the primordial power spectrum to the halo population and show the consequences of the spiky enhancement on the boost factor introduced in the previous section. We follow Ref.~\cite{Delos:2018ueo} (hereafter referred to as D18) and Ref.~\cite{Kawasaki:2021yek}, and we assume that $\mathcal{P}_{\mathcal{R}}(k) = \mathcal{P}_\star(k)$ (putting aside the smooth $\Lambda$CDM part of the primordial power spectrum). Assuming that halos are then formed isolated with a negligible rate of mergers we first caracterise their population with the BBKS formalism \cite{Bardeen:1985tr}. Then, we discuss the properties of the halos formed and the consequences for the boost-factor. In \refsec{sec:EST_HMF}, we will compare these results with an approach based on the EST.

\subsection{The population of peaks}

The BBKS formalism gives the average number of peak in the matter density field from the moments of the linear density contrast $\delta({\bf r}, z) = -1+\rho_{\rm m}({\bf r}, z)/ \overline{\rho}_{\rm m}(z)$ (with $\rho_{\rm m}$ the total matter density). The dimensionless linear matter power spectrum $\mathcal{P}_{\rm m}$ in the matter dominated era stems from the primordial power spectrum by introducing the transfer and the growth functions, $T(k)$ and $D(z)$ respectively. More precisely, we write \cite{Dodelson:2003ft}
\begin{equation}
    \mathcal{P}_{\rm m}(k, z) = \frac{4}{25}\left[ \frac{D(z) k^2 }{\Omega_{\rm m, 0} H_0^2} \right]^2 T^2(k) \mathcal{P}_{\mathcal{R}}(k) \, ,
    \label{eq:Pmatter}
\end{equation}
where $\Omega_{\rm m, 0}$ is the matter abundance and $H_0$ the Hubble rate today. The growth function accounts for the heightening of the density contrast (assuming that it remains linear) inside the horizon, and during the matter dominated era is simply given by $D(z)=(1+z)^{-1}$. The transfer function accounts for the fact that larger modes $k$ reentered the horizon earlier. In our analysis we have relied on the parametric formula found in Ref.~\cite{Eisenstein:1997ik}\footnote{We tried implementing both the zero baryon and non-zero baryon transfer functions from Ref.~\cite{Eisenstein:1997ik}. We have found that the latter generically leads to smaller boost factors as a consequence of the suppression on the matter power spectrum that is imparted when adding baryons. To allow for comparison with former works on UCMHs, we only show results considering the zero baryon transfer function. }. 

From the linear matter power spectrum, we introduce the weighted linear moments, smoothed on regions of size $R$,
\begin{equation}
    S_j(R) \equiv \int_0^{\infty} k^{2j} \mathcal{P}_{\rm m}(k, 0) |\hat{W}_R(k)|^2  {\rm d } \ln k \quad \forall j \in \mathbb{N}\, . 
    \label{eq:SmoothedVariance}
\end{equation}
In this expression, $W_R$ is the window function filtering the modes of scales smaller than $R$ and $\hat W_R$ is its Fourier transform. Several options for $W_R$ exist, this will prove irrelevant in this section but it will be addressed in greater details in \refsec{sec:EST_HMF}. Nonetheless, while the moments have been written in terms of a smoothing scale $R$ it is actually more convenient to think in terms of the mass of the overdensity instead of their physical size. We can assign a single mass to a given scale with the relation
\begin{equation}
    M \equiv \overline{\rho_{\rm m, 0}}\gamma_R R^{3} \, ,
    \label{eq:RtoM}
\end{equation}
where $\gamma_R$ is a numerical conversion factor. For a given choice of filter function, the value of $\gamma_R$ is not uniquely defined. The common prescription, that we follow in this work, is to require the normalisation condition on the volume of the structure, $\gamma_R\hat W_{R}(0)R^3 = 1$. Therefore, from now on we interchangeably write $R$ or $M$ to refer to an overdensity of size $R$ and associated mass $M$, assuming \refequ{eq:RtoM}. 

Halos form as initial overdensities grow to the point of gravitational instability and collapse. To define a linear threshold for this process, it is convenient to treat the problem from the linear regime and consider the linear extrapolation of the overdensities at the time of collapse. For spherical collapse in the matter-dominated era, the critical value for this extrapolated overdensity is $\delta_{\rm c} = 1.686$. The quantity of collapsed peaks / formed halos then depends on $\nu_{\rm c}(M, z)$, the ratio of $\delta_{\rm c}$ over the smoothed linear density variance. When working with linear perturbations, one can assume a fixed overdensity and a moving threshold barrier,
\begin{equation}
    \omega(z) = \delta_{\rm c} \frac{D(0)}{D(z)}\, , ~ \text{such that} ~ \nu_c(M, z) = \frac{\omega(z)}{\sqrt{S_0(M)}}\, .
\end{equation}
In full generality, the number density of collapsed peaks is then given by
\begin{equation}
    n = \frac{1}{(2\pi)^2 3^{3/2} } \left( \frac{S_2}{S_1}\right)^{3/2}\int_{\nu_{\rm c}}^{\infty} {\rm d} \nu \,   \int_0^{\infty} {\rm d} x \, f_{\rm BBKS}(x) \frac{e^{-\frac{(x-\nu\gamma)^2}{2(1-\gamma^2)}-\nu^2/2}}{ \sqrt{2\pi(1-\gamma^2)} }
    \label{eq:NumberOfCollapsedPeak}
\end{equation}
with $\gamma = S_1/(S_2 S_0)^{1/2}$ and $f_{\rm BBKS}$ defined in equation (A15) of Ref.~\cite{Bardeen:1985tr}. For the primordial power spectrum $\mathcal{P}_{\star}(k) = \As \ks \delta(k-\ks)$ the moments of the density field only depend on the mass through the window function and satisfy $S_j(M) \propto \ks^{3+2j}$. Thus $\gamma(M)=1$ and the exponential in \refequ{eq:NumberOfCollapsedPeak} reduces to a Dirac distribution. In D18 and Ref.~\cite{Kawasaki:2021yek}, they further consider that $|\hat W_{R}(\ks)| = 1$, which yields $\nu_{\rm c}(M, z) = \nu_{\rm c}(z)$ independent of the mass. The number density of collapsed peaks is thus simply
\begin{equation}
    n(z) = \frac{\ks^3}{(2\pi)^2 3^{3/2}}\int_{\nu_{\rm c}(z)}^{\infty} {\rm d} \nu  \, e^{-\nu^2/2} f_{\rm BBKS}(\nu)
\end{equation}
In this picture and under the assumption that halos form isolated and do not merge into larger structures, the number density of halos per mass and formation redshift can be written,
\begin{equation}
    \frac{\partial^2 n(M, z_{\rm f}\, | \, z)}{\partial  M \partial z_{\rm f} } \equiv \Theta(z_{\rm f} -z) \delta\left[M-m(z, z_{\rm f})\right] \left. \frac{{\rm d} n}{{\rm d} z} \right|_{z = z_{\rm f}} \ .
    \label{eq:mass_function_BBKS}
\end{equation}
The virial mass at redshift $z$ is fixed by the formation redshift (corresponding to the collapse time) to $m(z, z_{\rm f})$ given below, in \refsec{sec:BBKS_HMF_Properties}. The Heaviside function ensures that, at redshift $z$, only halos with a larger formation redshift are populating the Universe. 

\subsection{The density profile of UCMHs}
\label{sec:BBKS_HMF_Properties}

 In the D18 model they relate, from fits on their own numerical simulations, the profile to the formation redshift. They have shown that UCMHs form with a Moore profile  \cite{Moore:1999gc}, $\gamma=3/2$, with the parameters,
\begin{equation}
    \begin{cases}
    r_{\rm s} & =f_1 \ks^{-1}(1+z_{\rm f})^{-1} \\
    \rho_{\rm s} & = f_2 \overline{\rho_{\rm m,0}}(1 + z_{\rm f})^3 
    \end{cases}
\label{eq:MooreParameters2}
\end{equation}
where $f_1 \simeq 0.7$ and $f_2 \simeq 30$. We recall that $\ks$ is the scale of the additional spike in the primordial power spectrum. Combining \refequ{eq:MooreParameters2} with \refequ{eq:definition_virial_parameters}, we arrive at the following expressions for the concentration and mass of UCMHs
\begin{equation}
\begin{cases}
\displaystyle \frac{c^3}{\mu(c)} = 3f_2\frac{\overline{\rho_{\rm m, 0}}(1+z_{\rm f})^3}{\Delta \rho_{\rm c}(z)} \\[10pt]
 \displaystyle m =  \overline \rho_{\rm m, 0} f_1^3 f_2 \mu(c) \ks^{-3} \, .
\end{cases}
\label{eq:MooreParameters}
\end{equation}
Notice that the first of the previous two equations defines implicitly the concentration as a function $c = c(z, z_{\rm f})$. Note also that introducing these mass and concentration parameters can be seen as superfluous since the D18 model directly gives the halo properties in terms of the scale radius and density, however, this will allow us to connect to our formalism described in the next section.

\section{Halos in the EST formalism}
\label{sec:EST_HMF}

In this section we depart from the BBKS formalism to give a complementary approach to D18 using the EST/Press-Schecheter formalism \cite{Press:1973iz, Bond:1990iw}. We show that despite some well identified drawbacks this methods allows us to better describe the halo population at late time and quantify the theoretical uncertainties associated to it. In regard to \refequ{eq:mass_function_BBKS} here we write
\begin{equation}
    \frac{\partial^2 n(M, z_{\rm f}\, | \, z)}{\partial  M \partial z_{\rm f} } \equiv \frac{{\rm d}n(M\, | \, z)}{{\rm d} M} p_{z_{\rm f}}(z_{\rm f}\,  |\, M, z)  \, ,
\end{equation}
where ${\rm d} n / {\rm d}M$ is the cosmological mass function and $p_{\rm z_{\rm f}}$ is the probability distribution of formation redshift for a halo of mass $M$ roaming the Universe at redshift $z$. We discuss both quantities in the following.

\subsection{The cosmological halo mass function}

As detailed above, in D18 the mass function of UCMHs is evaluated using the BBKS formalism. Their numerical simulations show that UCMHs should form early and isolated (with mergers being extremely rare). Therefore the subsequent UCMH mass function obtained from the BBKS formalism should be safe from the cloud-in-cloud shortcomings. In other words, it is always possible to associate a peak of a given scale to a UCMH with a certain mass. Nonetheless, in order to properly evaluate the boost factor between the matter radiation equality and today, we need to describe a population made of a mixture of both UCMHs and \emph{standard} halos (formed from successive mergers) with a flatter inner density profile, $\gamma < 3/2$ in \refequ{eq:generlized_NFW}. As mentioned in the introduction, we discard UCMHs surviving late-time accretion and mergers, as well as the presence of ultracompact subhalos which would further add a subhalo boost to the boost. Thus, in the following we consider stadard halos with NFW profiles \cite{Navarro:1995iw} corresponding to $\gamma=1$. For these latter ones, the BBKS formalism fails, as the aforementioned identification between peaks and halos breaks down. We derive here the halo mass function from the EST/Press-Schechter formalism.

The computation starts from the zeroth-order smoothed variance as defined in \refequ{eq:SmoothedVariance}. Due to the additional spike in the curvature power spectrum, it can be written as the sum 
\begin{equation}
    S_0(R) = S_{\rm PL}(R) + S_\star(R) \, ,
 \end{equation}  
 where,
 \begin{equation}
 \begin{cases}
     \displaystyle S_{\rm PL}(R) = \frac{4}{25}\left[ \frac{ D(0) k_0^2}{\Omega_{\rm m, 0} H_0^2}\right]^2 \mathcal{A}_{\rm s}  \int_0^{\infty} T^2(k) \left(\frac{k}{k_0}\right)^{n_s+3}  |\hat W_R(k)|^2 {\rm d} \ln k \\[12pt] 
     \displaystyle S_\star(R) = \left\{ \Sigma_\star \equiv \frac{4}{25}\left[ \frac{ D(0) k_0^2 }{\Omega_{\rm m, 0} H_0^2}\right]^2 \As T^2(\ks)\left(\frac{\ks}{k_0}\right)^{4}\right\} |\hat W_R(\ks)|^2 \, .
     \end{cases}
     \label{eq:VarianceWithSpike}
\end{equation}

Instead of the usual real-space top-hat window $W_R(r) = 3/(4\pi R^3)\Theta(R-r)$ -- with $\Theta$ the Heaviside distribution --, in the following we work with the sharp $k$-space filter window function $ W_R(k) = \Theta(1-kR)$ defined in the Fourier space directly. Indeed, the latter is the only natural choice in the EST framework\footnote{Using other window functions in the excursion set theory is extremely challenging, see for instance Ref.~\cite{verechtchaguinaFirstPassageTime2006}.}. Moreover, when the power spectrum is highly dependant on $k$ (as when it exhibits a peak) the mass function behaviour can become erroneously driven by the window function with other filters \cite{Schneider:2013ria, Schneider:2014rda}. The main drawback associated to the sharp-$k$ window lies in the definition of the $\gamma_R$ factor to relate mass and scale that is not physically meaningful. Nonetheless, choosing $\gamma_R = 6\pi^2$ from our normalisation choice $\gamma_R\hat W_{R}(0)R^3 = 1$,  has been proven to be in good agreement with simulations \cite{Lacey:1994su}\footnote{Although it remains somewhat dependant on the halo definition and on the halo finding method.}. 

\begin{figure}[t!]
\centering
\includegraphics[width=0.43\linewidth]{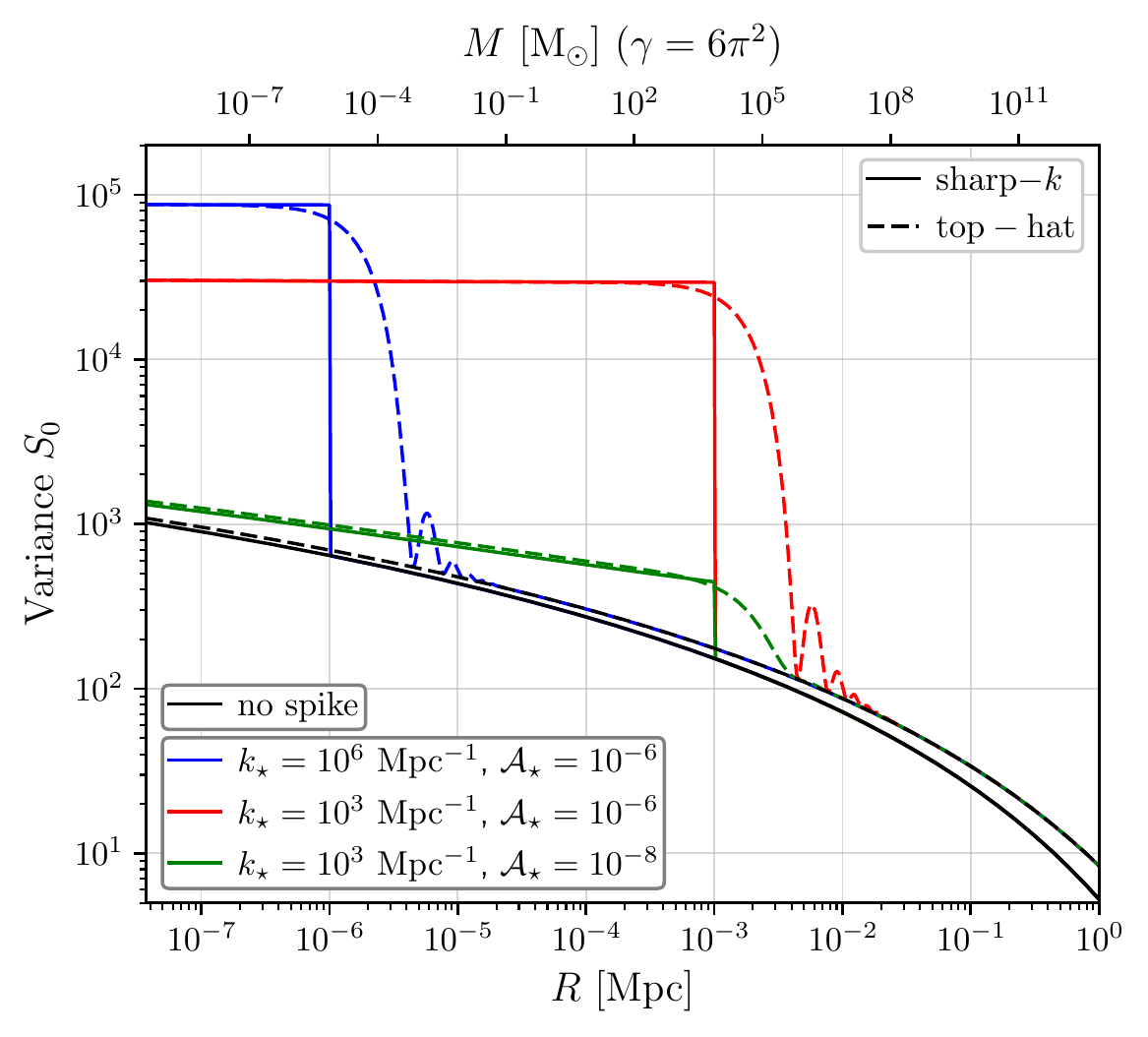} \hfill
\includegraphics[width=0.54\linewidth]{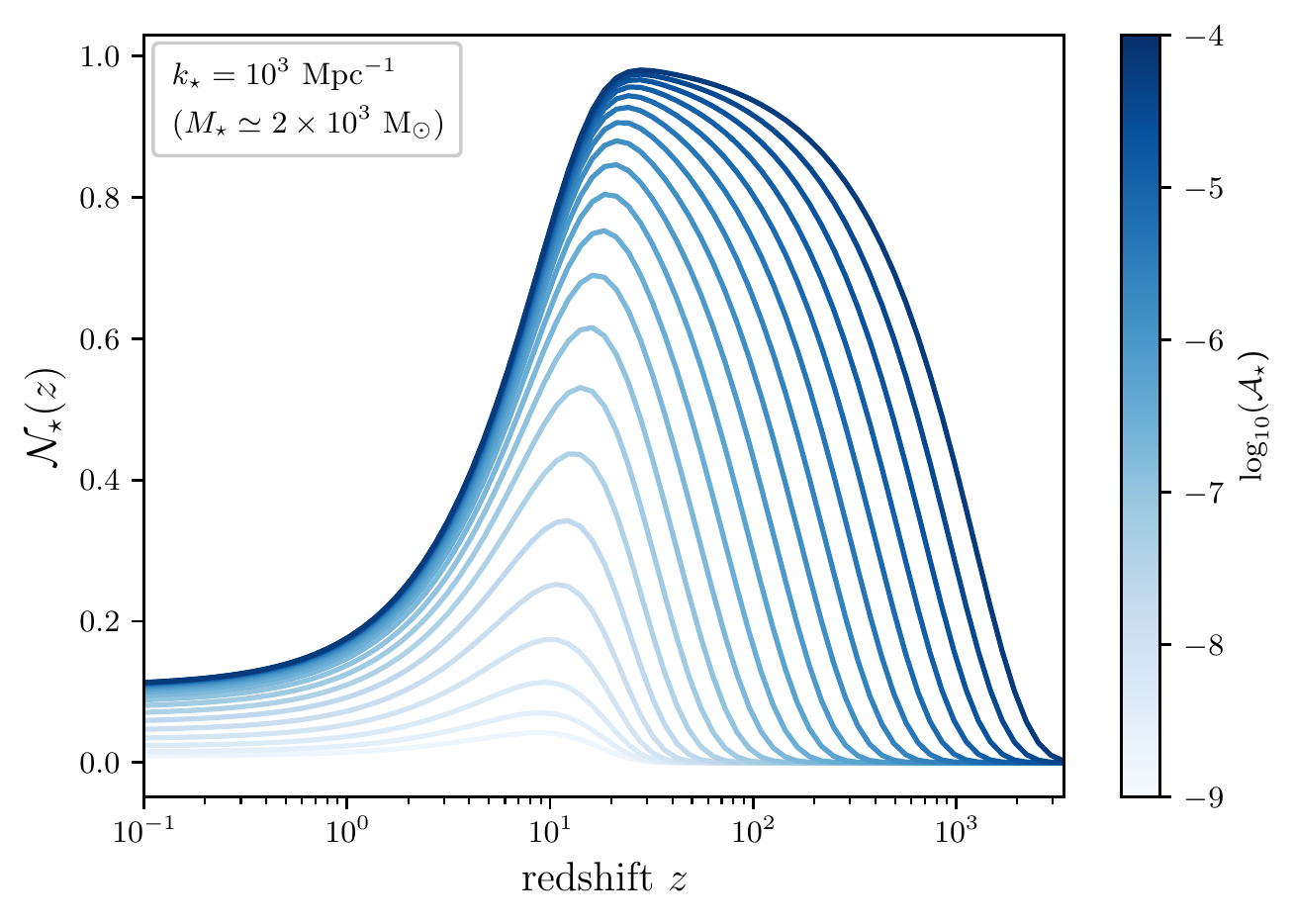}
\caption{ {\bf Left panel.} Variance of the matter power spectrum for different spike amplitude and position with respect to the radius (and mass). The solid line show the result from the sharp-$k$ window while the dashed lines correspond to the real-space top-hat. The correspondence between radius and mass is performed assuming $\gamma = 6\pi^2$. {\bf Right panel.} Mass fraction contained in the spike of the mass function at $k_\star = 10^3$~$\rm Mpc^{-1}$ ($M_\star = 2.3 \times 10^3$ M$_\odot$) in terms of the redshift $z$. The blue color scale covers a range of amplitudes $\mathcal{A}_\star$, logarithmically distributed from $10^{-9}$ to $10^{-4}$.}
\label{fig:VarianceMatterPowerSpectrum}
\end{figure}

In the left panel of \reffig{fig:VarianceMatterPowerSpectrum} we show $S_0$ in terms of the radius (and mass) for the real-space top-hat (dashed lines) and sharp-$k$ (solid lines) window functions. Choosing a sharp-$k$ cutoff results in a discontinuous drop at $R = R_\star =1/\ks$ because of the Heaviside distribution in Fourier space. Different scenarios (\emph{i.e.}, different values of $\As$ and $\ks$) are also represented in three different colors. To get a sense of the scaling of the variance with the different parameters, the transfer function of Ref.~\cite{Eisenstein:1997ik} can be approximated at the subpercent level on the range $[1, 10^{10}]~{\rm Mpc^{-1}}$ by the expression -- see equation (4.240) in Ref.~\cite{moGalaxyFormationEvolution2010},
\begin{equation}
    T(k) \simeq c_{\rm T} \left(\frac{k}{k_{\rm T}}\right)^{-2} \ln\left(\frac{k}{k_{\rm T}}\right)
\end{equation}
with $c_{\rm T}= 0.23$ and $k_{\rm T} = 6.2 \times 10^{-2} ~{\rm Mpc^{-1}}$. Plugging numbers into \refequ{eq:VarianceWithSpike} then gives the simple relation, showing that $\Sigma_\star$ depends logarithmically on $k_\star$ (or equivalently $R_\star$),
\begin{equation}
    \Sigma_{\star} \simeq \frac{\mathcal{A}_\star}{3.2\times 10^{-9}} \ln^{2}\left(R_\star{k}_{\rm T}\right) \, .
    \label{eq:SigmaStarApprox}
\end{equation}
For $\mathcal{A}_\star = 10^{-6}$ (blue and red curves), $S_0(R < R_\star) \simeq \Sigma_\star$. The height of the plateaus are thus well evaluated by \refequ{eq:SigmaStarApprox}.  Between $k = 10^{3}~{\rm Mpc^{-1}}$ (red) and $k = 10^{6}~{\rm Mpc^{-1}}$ (blue) the increase is of order $\sim 3$ corresponding to the ratio of squared logarithms. Moreover, changing $\mathcal{A}_\star$ from $10^{-8}$ to $10^{-6}$ (green and red curves) does increases the variance on small scales by two orders of magnitude (although for $\mathcal{A_\star}=10^{-8}$, $S_{\rm PL}$ is not negligible with respect to $\Sigma_\star$). \\

In the EST framework, the halo mass function is given by
\begin{equation}
    \frac{{\rm d} n(M \, | \, z)}{{\rm d} M} = \frac{\overline \rho_{\rm m,0}}{M} \frac{\nu(M, z)}{2S_0(M)} \left| \frac{{\rm d}S_0}{{\rm d} M} \right| f_{\rm EST}\left(\nu(M, z)\right) \, ,
\end{equation}
and in the spherical collapse framework, the function $f_{\rm EST}$ takes the same form than that of Press-Schechter
\begin{equation}
    f_{\rm EST}(\nu) \equiv \sqrt{\frac{2}{\pi}}e^{-\nu^2/2} \, .
\end{equation}
In Refs.~\cite{Sheth:1999mn, Sheth:1999su} the authors give a mass function in better agreement to simulations results by considering ellipsoidal collapse. Consequently, they introduce a collapse barrier that depends on $S_0$ (and not only on $z$) and a different form for the function $f$. However, this approach relies on fits that may not apply to this much differently shaped power spectrum. Hence, we choose to work here with the Press-Schechter formula. Nonetheless, because of the Heaviside distribution in $S_\star(R)$, the expression of the mass function is not straightforward. With a proper regularisation method, and to ensure a correct normalisation as shown in \refapp{app:HeavisideInPS}, we find the expression
\begin{equation}
    \frac{{\rm d} n(M \, | \, z)}{{\rm d} M} = \frac{\overline \rho_{\rm m,0}}{M} \left\{ \frac{\nu(M, z)}{2 S_0(M)} \left| \frac{{\rm d} S_{\rm PL}}{{\rm d} M} \right|f\left(\nu(M, z)\right)  +  \delta(M- M_\star) \mathcal{N}_\star(z) \right\} \, .
    \label{eq:full_mass_function}
\end{equation}
The mass fraction of halos \emph{from} the spike at $M_\star$ is
\begin{equation}
    \mathcal{N}_\star(z) \equiv \int_{\nu_{-}(z)}^{\nu_{+}(z)} f(\nu) {\rm d} \nu  = {\rm erf}\left(\frac{\nu_+(z)}{\sqrt{2}} \right) - {\rm erf}\left(\frac{\nu_-(z)}{\sqrt{2}} \right)\, .
\end{equation}
where the integral boundaries are
\begin{equation}
    \nu_{-}(z) \equiv  \frac{\omega(z)}{\sqrt{S_{\rm PL}(M_\star) +  \Sigma_\star}} \quad {\rm and} \quad 
    \nu_{+}(z)  \equiv \frac{\omega(z)}{\sqrt{S_{\rm PL}(M_\star)}} \, .
\end{equation}
The fraction $\mathcal{N}_\star$ is shown in the right panel of \reffig{fig:VarianceMatterPowerSpectrum} in terms of the redshift and $\Sigma_\star$. The halos of mass $M_\star$ only dominate on a specific redshift range provided that the amplitude of the spike in the power spectrum is large enough. Indeed, at low redshifts, all the halos formed with mass $M_\star$ gradually become larger by accretion or they merge to become subhalo of larger structures. Therefore we conservatively assume that they loose their cuspy profile and adopt an NFW profile when they reach larger masses or that they are partially tidally stripped subhalos that we neglect.

\subsection{The distribution of formation redshift}
\label{sec:distrib_formation_redshift}

The probability distribution function of formation redshift, $p_{z_{\rm f}}$, is the most challenging quantity to define in the EST. Indeed, the formation redshift is not a well defined quantity for halo forming through accretion and successive mergers. Nonetheless, it is necessary to estimate it in order to get to the halo properties (such as the concentration). A convenient approximation is to assume that the formation redshift is directly set by the variance of the matter power spectrum according to $S_0(M) = \omega^2(z_{\rm f})$. Let us call $Z_{\rm f}(M, z)$ the solution of this equation. It can be expressed as
\begin{equation}
    Z_{\rm f}(M) \equiv  D^{-1}\left( \frac{\delta_c D(0)}{\sqrt{S_0(M)}} \right)
    \label{eq:Znaive}
\end{equation}
with $D^{-1}$ the inverse of the growth function $D$. This value is a constant of $z$ which means, $Z_{\rm f}$ can be lower than $z$ for the largest masses. In practice, however, the corresponding populations of halos is suppressed in the mass function. Therefore, we should see this definition as an estimate of the statistical average upon all halos. In that scenario, the probability distribution function for $z_{\rm f}$ simply reduces to
\begin{equation}
    p_{z_{\rm f}}(z_{\rm f} \, | \, M, z)  = \delta \left( z_{\rm f} - Z_{\rm f}(M) \right) \, .
\end{equation}
However, owing to the discontinuity of $S_\star$, for $M = M_\star$ this definition is ambiguous\footnote{One could argue that such a discontinuity is only due to the choice of a sharp $k$-space window function and that another prescription would give a natural regularisation. Nevertheless, the relationship between formation redshift and mass would then be fully set by the shape of the window function, which is not physical.} in our case.  We can introduce several approximations for the value of $Z_{\rm f}(M_\star, z)$ and in particular an average value given by
\begin{equation}
\begin{split}
    \left< Z_{\rm f}(M_\star) \right>(z)   \equiv \frac{1}{\mathcal{N}_\star(z)} \int_{\nu_{-}(z)}^{\nu_{+}(z)}  D^{-1}\left(\nu D(z) \right) f(\nu) {\rm d} \nu 
    \end{split}
    \label{eq:Zfaverage}
\end{equation}
which is bounded by
\begin{equation}
    Z_{\rm f}(M_\star^{\pm})  \equiv \lim_{M\to M_\star^{\pm}} Z_{\rm f}(M) \,
    \label{eq:Zfaverage2}.
\end{equation}
These bounds allow us to bracket the theoretical uncertainties inherent to the choice of formation redshift definition, as we will discuss in \refsec{sec:boost_refined_model}. More details on the formation redshift are given in \refapp{app:FormationRedshift}.

\subsection{Properties of the halos}
\label{sec:EST_HMF_Properties}

Thanks to this framework, we can describe a population made of a mixture of both UCMHs and standard halos. Indeed, looking at the left panel of \reffig{fig:VarianceMatterPowerSpectrum}, we see there is a natural way to split the mass interval: we consider that light halos with $m \le M_\star$ (where $S_0 \simeq S_\star$) are UCMHs and form Moore profiles, whereas larger halos with $m > M_\star$ (where $S_0 = S_{\Lambda \mathrm{CDM}}$) stem from successive accretions resulting in an NFW profile. \\

{\bf \emph{Standard} NFW halos}. To evaluate the properties of these halos we rely on Ref.~\cite{Maccio:2008pcd}. According to Ref.~\cite{Diemer:2018vmz}, they form in two steps, starting by a phase of rapid accretion until the halo reaches a concentration $c\sim 4$ after which they \emph{pseudo-evolve}. To capture this non-trivial evolution history, the model introduces a modified formation redshift $\tilde z_{\rm f}$ satisfying $S_{\rm PL}(FM) = \omega^2(\tilde z_{\rm f})$ (once again here this new formation redshift is not to be taken as an exact physical quantity but rather like an estimate and a mathematical intermediary to get to the concentration). The parameter $F$ is a free parameter calibrated on simulations to $F=0.01$. The pseudo-evolution is then characterised, not by a constant scale density but by a constant \emph{core density} $\tilde \rho_{\rm s} \equiv \mu(c)\rho_{\rm s}$. In conclusion, for the NFW halos, the virial mass and concentration are set by
\begin{equation}
\begin{cases}
     \displaystyle c^3(z, z_{\rm f}) = \frac{\Delta K^3}{\Omega_{\rm m}(\tilde z_{\rm f})} \frac{\overline{\rho}_{\rm m, 0}(1+ \tilde z_{\rm f})^3}{\Delta \rho_{\rm c}(z)} \\[10pt]
    \displaystyle m= M
    \end{cases}
        \label{eq:concentrationNFW}
\end{equation}
where $K = [3\tilde\rho_{\rm s}/(\Delta \rho_{\rm c})]^{1/3} = 4.2$ is also calibrated on simulations and assumed independent of $\tilde z_{\rm f}$. We can compare the expression of the concentration in \refequ{eq:concentrationNFW} to that of \refequ{eq:MooreParameters} obtained from the scale parameters derived in D18. Both have the same structure as they derive from similar arguments. The small difference appears due to the accretion phase of NFW halos resulting in a constant core density (instead of constant scale density). Moreover note here that, because we are working in the EST approach, the virial mass is directly set as the one entering in the expression of the mass function. \\

{\bf UCMHs (with a Moore profile)}. We consider that they form at redshift $Z_{\rm f}(M)$ and, using the fits on numerical simulations obtained in D18, they have a concentration and virial mass defined by \refequ{eq:MooreParameters}. As previously mentioned, in this framework, the formation redshift is only defined statistically, therefore it yields $z_{\rm f} < z$ for an exponentially suppressed population of halos. This has consequences for the regularisation of the central part of the Moore profile  in the $s$-wave scenario -- \emph{c.f.}, \refequ{eq:regularisation_density_profile}. Indeed, configurations where $t(z) \le t(z_{\rm f})$ and $\rho_{\rm max}$ negative or infinite are  allowed. This issue is solved by introducing a time-difference regulator, $\Delta t_{\rm vir}$, to avoid divergences when $z_{\rm f} \simeq z$. The latter is defined as the typical elapsed time between turnaround and virisalisation of halos in the spherical collapse model. The notion of a virialised halo itself is unclear on shorter timescales. More precisely we have, in the matter dominated era,
\begin{equation}
    \Delta t_{\rm vir}(z_{\rm f}) = \frac{1}{2}t(z_{\rm f}) 
\end{equation}
and we trade \refequ{eq:regularisation_density_profile} for
\begin{equation}
    \rho_{\rm max} = \frac{m_\chi}{\sigma_0 {\rm max}\left\{[t(z) - t(z_{\rm f})], \Delta t_{\rm vir}\right\}}
    \label{eq:rho_maxx}
\end{equation}
We emphasize again that owing to the exponential suppression of the concerned halo population the chosen regularisation procedure does not have a strong impact and is mostly here to avoid numerical divergences and non-positive densities.

\subsection{Boost factor: a refined model}
\label{sec:boost_refined_model}

\begin{figure}[t!]
\centering
\includegraphics[width=0.49\linewidth]{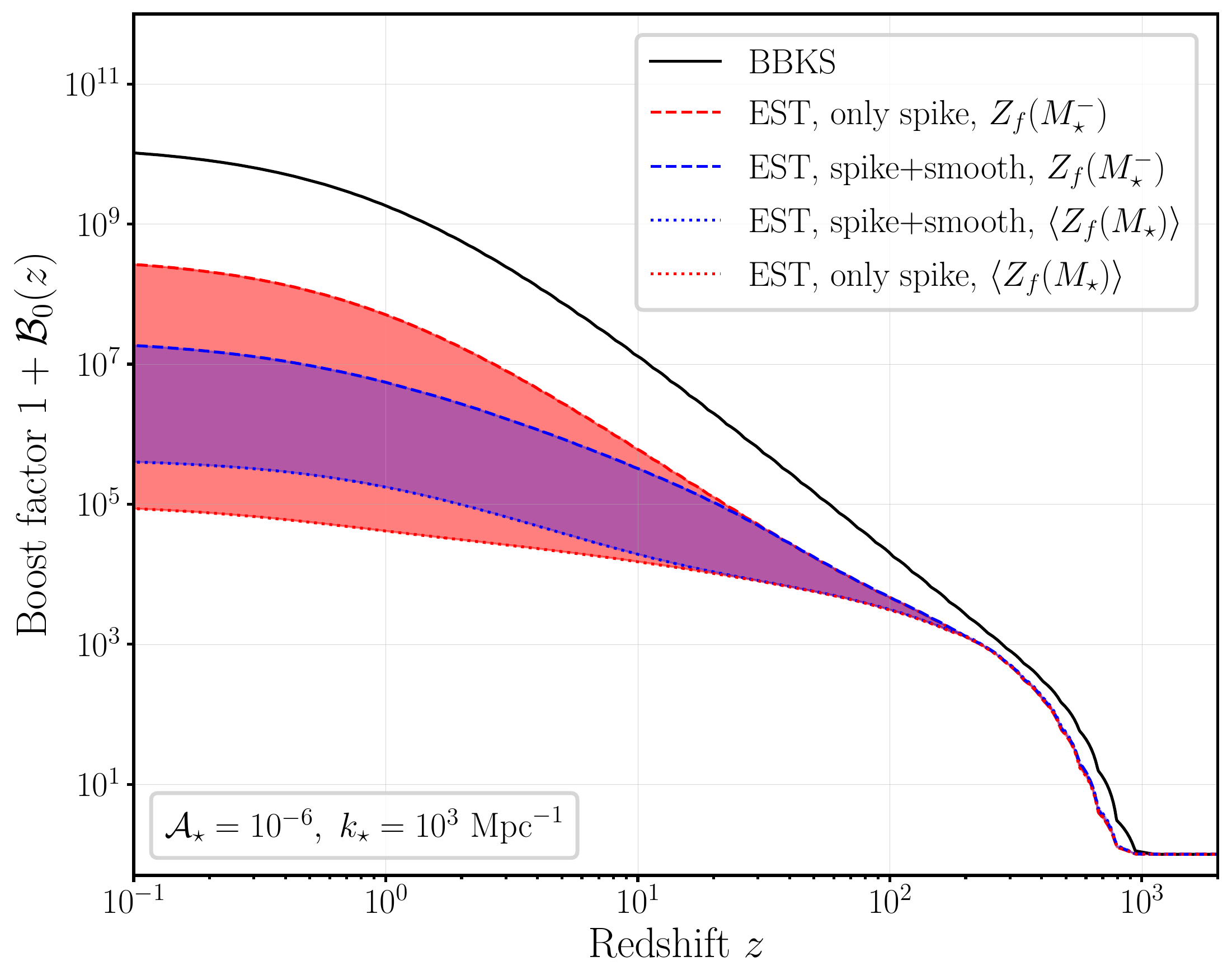} 
\includegraphics[width=0.49\linewidth]{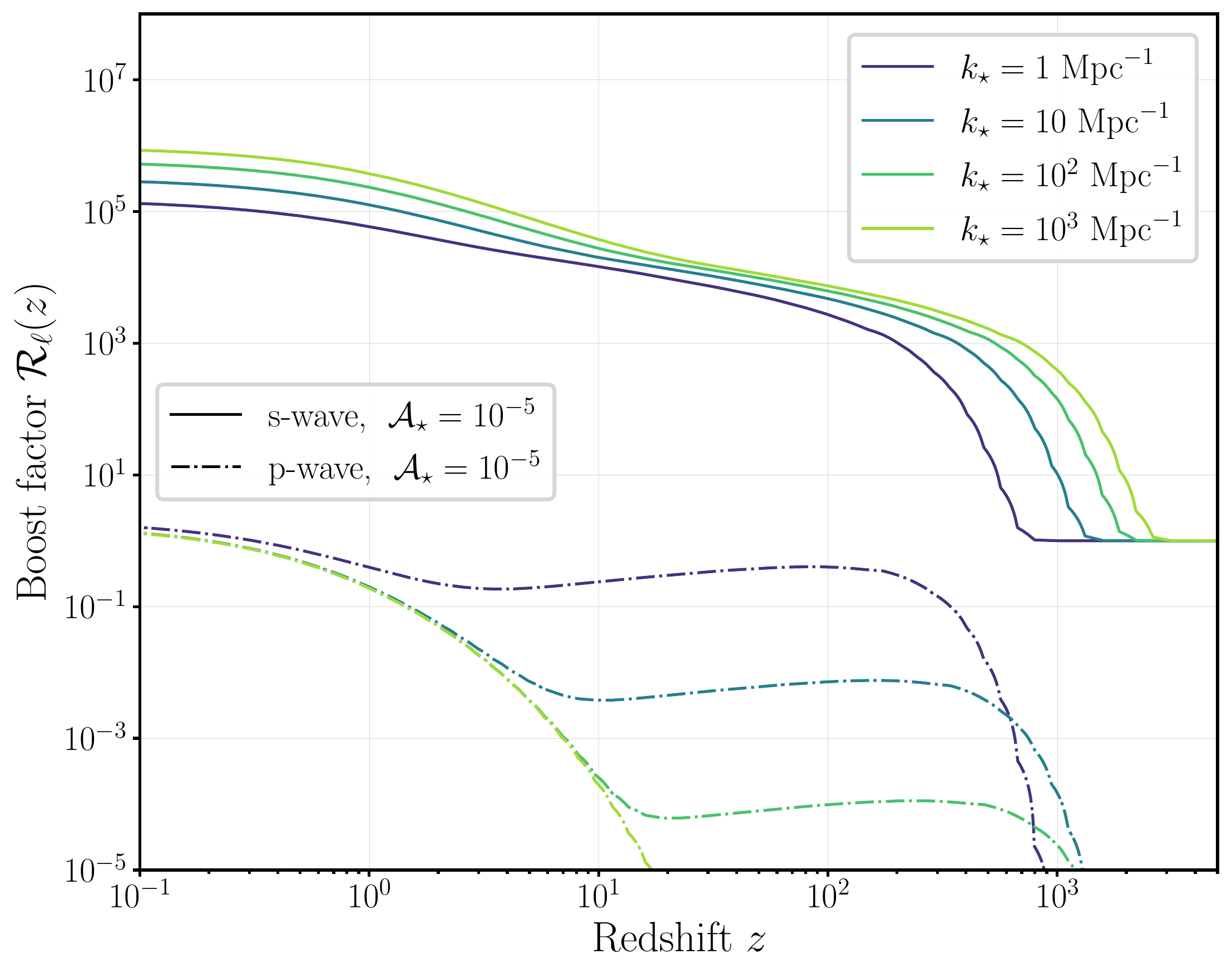} 
\caption{{\bf Left panel:} Boost factor for $s$-wave DM annihilations, computed 
under the EST and BBKS formalisms. These curves consider a spike in the primordial spectrum with $\mathcal{A}_{\star} =10^{-6}$ and  $k_{\star} = 10^3 \ \rm{Mpc}^{-1}$. For EST, we have considered four different prescriptions that allow to bracket the theoretical uncertainty: two choices for the formation redshift (either $\left< Z_{\rm f}(M_\star) \right>$ or $Z_{\rm f}(M_\star^{-})$, see \refequ{eq:Zfaverage}-\refequ{eq:Zfaverage2}) and the addition or not of the smooth $\Lambda$CDM component in the primordial power spectrum. {\bf Right panel:} Boost factor for $s$- and $p$-wave DM annihilations, for a fixed amplitude of the spike $\As = 10^{-5}$, and several values of $k_{\star}$. These curves have been computed using the EST formalism with the prescription ``spike+smooth, $\left< Z_{\rm f}(M_\star) \right>$'' (blue dotted line in the left panel). }
\label{fig:boost_several_prescriptions}
\end{figure}

We have now given all the ingredients to calculate the boost factor in the presence of a spike in $\mathcal{P}_{\mathcal{R}}(k)$, using both the BBKS and the EST formalisms. With our improved EST treatment, we have shown for the first time how to account for a mixed population of halos (consisting of UCMHs with Moore profiles and standard halos with NFW profiles) and how to incorporate $p$-wave DM annihilations. Those novel theoretical results are summarized in \reffig{fig:boost_several_prescriptions}. In the left panel we compare the predictions from BBKS and EST, and use the latter to estimate the theoretical uncertainties associated to the boost factor. In particular, we consider four different prescriptions, coming from two different choices of the formation redshift and the addition or not of the smooth power-law $\Lambda$CDM component in $\mathcal{P}_{\mathcal{R}}(k)$. Comparing the dashed and dotted lines, we see that different formation time choice yields orders of magnitude differences at small redshifts. These differences are reassuringly decreased when including the smooth $\Lambda$CDM component in the primordial spectrum (compare blue and red bands). Nevertheless, we remark that these differences do not impact $z > 10^2$ where the CMB is most sensitive to (although they are expected to be important for late time probes like the 21 cm signal)\footnote{Throughout this paper, and unless stated otherwise, every time we show results using EST, we are implicitly considering the prescription ``spike+smooth, $\left< Z_{\rm f}(M_\star) \right>$'', which we regard as the most physical. }. In the right panel of \reffig{fig:boost_several_prescriptions} we compare the boost factor for $s$- and $p$-wave DM annihilations, for a fixed amplitude of the spike $\As$ and several wavenumbers $\ks$. We see that the $p$-wave boost is smaller than the $s$-wave one by at least four orders of magnitude. In addition, the $p$-wave boost sharply decreases for large values of $\ks$ (contrarily to the $s$-wave boost), which was to be expected given the scalings of the 1-halo boosts $\mathcal{B}_{\ell, h}$ with $\ks$ that we discussed at the end of \refsec{sec:one_halo_boost}. In spite of this, we will show in \refsec{sec:results_pwave} that $p$-wave annihilation can still lead to competitive bounds on the primordial spectrum at small $\ks$ for cross-sections not yet ruled out by the strongest astrophysical constraints.

\section{Results}
\label{sec:results}

To derive the 95 $\%$ CL bounds on the amplitude $\As$ and wavenumber $\ks$ of the spike in the primordial spectrum, we perform comprehensive MCMC analyses using the public code {\tt MontePython-v3} \cite{Brinckmann:2018cvx} interfaced with our modified version of {\tt ExoCLASS} \cite{Stocker:2018avm}. We carry out the analyses with a Metropolis-Hasting algorithm, assuming flat priors on the $\Lambda$CDM parameters $\{ \omega_b, \omega_{\mathrm{cdm}}, H_0, \ln{\left( 10^{10}A_s \right)}, n_s, \tau_{\rm reio} \}$ and logarithmic priors on $\ks$ and $\As$, namely,
\begin{align*}
\ 0 \ \leq \ \mathrm{Log}_{10} ( \ks/ &\mathrm{Mpc}^{-1}) \ \leq 7, \\
 -8 \ \leq  \ \mathrm{Log}_{10} &\As  \ \leq  \ -4.
\end{align*}
We remark that those priors are largely uninformative and the bounds are physically well-motivated. In particular, very small spike amplitudes $\As < 10^{-8}$ leave the CMB spectra essentially unaffected, whereas very large spike amplitudes $\As > 10^{-4}$ can either produce UCMHs during the radiation era or even lead to a non-negligible population of PBHs, hence breaking some of the assumptions on which the formalism presented here is built (see \refapp{app:prior_region} for more details on this issue). In a similar vein, for very small wavenumbers $\ks < 1 \  \mathrm{Mpc}^{-1}$, the shape of the primordial spectrum is strongly constrained by CMB and Lyman-$\alpha$ observations, whereas for very large wavenumbers $\ks > 10^7 \  \mathrm{Mpc}^{-1}$ we don't expect UCMHs to form since this corresponds to scales smaller than the free-streaming length of typical WIMP particles \cite{Bringmann:2009vf} \footnote{See \refapp{app:k_fs} for details on how the addition of a free-streaming scale $k_{\rm fs}$ impacts the constraints. }.\

We adopt the \Planck~collaboration convention in modelling free-streaming neutrinos as two massless species and one massive with $m_\nu = 0.06 \ \mathrm{eV}$. For all the analyses that we present here, we include the following data-sets:
\begin{itemize}
    \item The low-$\ell$ CMB TT,EE and high-$\ell$ TT, TE, EE data + the lensing potential reconstruction from \Planck~2018 \cite{Planck:2018vyg}. We make use of a Cholesky decomposition as implemented in {\tt MontePython-v3} to handle the large number of nuisance parameters \cite{Lewis:2013hha}. 
    \item The Baryon Acoustic Oscillation (\BAO) measurements from 6dFGS at $z=0.106$ \cite{Beutler:2011hx}, SDSS DR7 at $z=0.15$ \cite{Ross:2014qpa} and BOSS DR12 at $z=0.38, 0.51$ and $0.61$ \cite{BOSS:2016wmc}.
    \item The \Pantheon~catalogue of SuperNovae of type Ia (SNIa), spanning redshifts $ 0.01 < z < 2.3$ \cite{Pan-STARRS1:2017jku}. We marginalize over the nuisance parameter $\mathcal{M}$ describing the SNIa calibration. 
\end{itemize} 
We note that the presence of UCMHs has a direct impact only on the TT, TE, and EE CMB spectra (through the modification in the thermal history of the universe), but the addition of CMB lensing, BAO and SNIa data is still useful in order to break parameter degeneracies. To bracket the uncertainties in the $\As$ vs. $\ks$ constraints, in the following we present results under various different assumptions: the velocity dependence of the cross section ($s$- or $p$-wave), the formalism used to compute the boost (BBKS or EST), as well as different DM masses and annihilation channels. We assume MCMC chains to be converged using the Gelman-Rubin criterion $R-1 < 0.01$ \cite{Gelman:1992zz}.

\subsection{The $s$-wave case }

\subsubsection{Comparison between EST and BBKS}

Let us focus first on the case of $s$-wave DM annihilations. To start, we would like to compare the constraints obtained using either the BBKS formalism (applied only to the spiky component in $\mathcal{P}_{\mathcal{R}}(k)$, see \refsec{sec:BBKS}) or the EST formalism (applied to both the smooth $\Lambda$CDM and spiky components of $\mathcal{P}_{\mathcal{R}}(k)$, see \refsec{sec:EST_HMF}). In this subsection, we consider DM annihilating into $b\bar{b}$ with a mass $m_{\chi} =1 \ \rm{TeV}$ and a thermal relic cross-section $\sigma_0 = 3\times 10^{-26}~\rm cm^3 s^{-1}$. \

\begin{figure}[t!]
\centering
\includegraphics[width=0.7\linewidth]{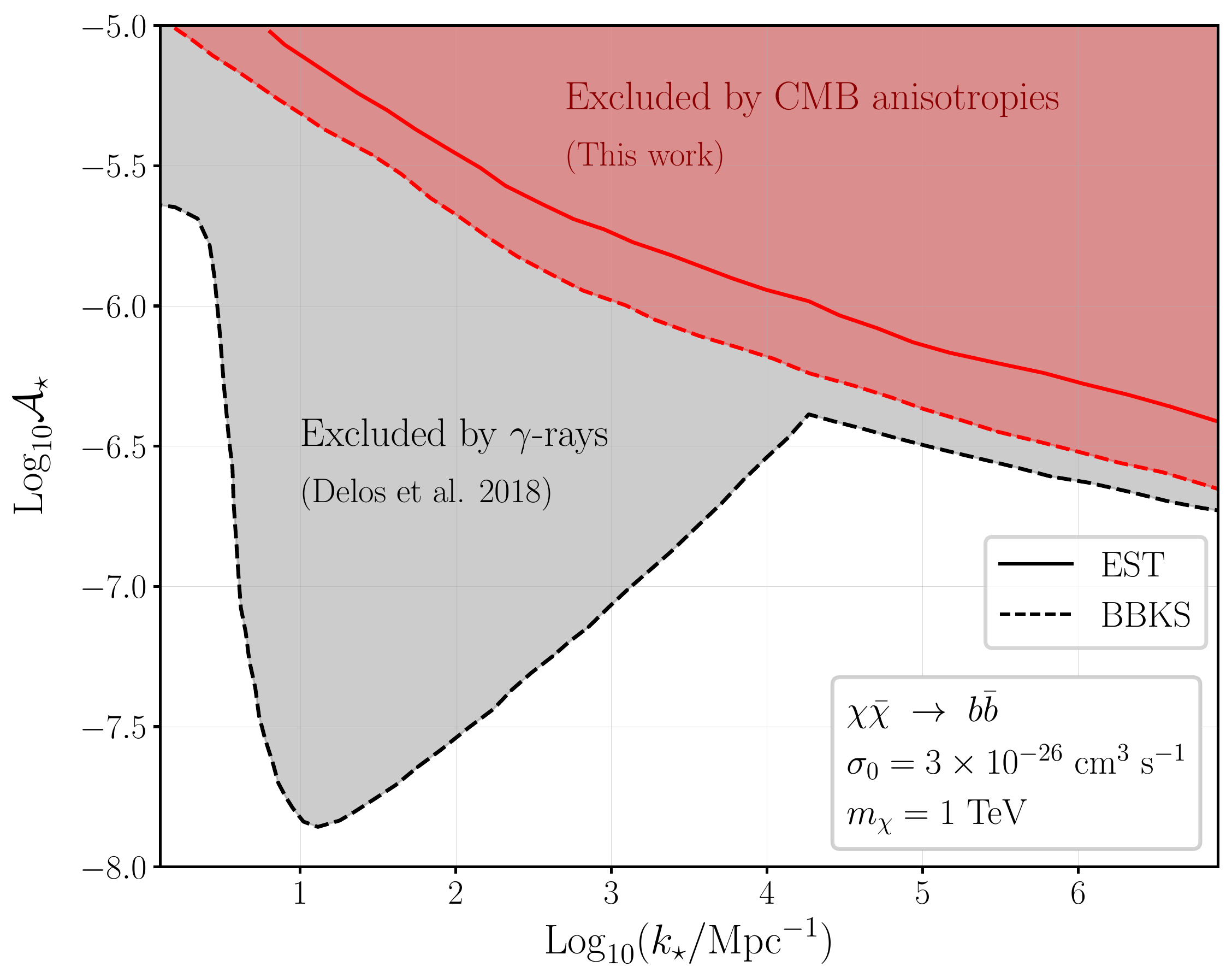} 
\caption{95\% C.L. bounds from \Planck+\BAO+\Pantheon~on the amplitude of the spike in the primordial spectrum as a function of its wavenumber, using either EST or BBKS formalism for the computation of the boost factor. For comparison, we also show the $\gamma$-ray constraints from \cite{Delos:2018ueo} (using both point-sources and the diffuse flux), which were derived with the BBKS formalism. All curves assume that DM annihilates into $b\bar{b}$ with a mass $m_{\chi} =1 \ \rm{TeV}$ and a $s$-wave thermal relic cross-section.}
\label{fig:constraints_est_vs_bbks}
\end{figure}

In \reffig{fig:constraints_est_vs_bbks} we show the 95 \% C.L. excluded region in the $\As$ vs. $\ks$ plane from a \Planck+\BAO+\Pantheon~analysis,  derived using both the EST and the BBKS formalism. We see that the constraints in the BBKS case are stronger than in the EST case, but not by a significant amount. This was to be expected; even if BBKS generically predicts larger boosts than EST, the differences at $z \sim 10^3$ (where the CMB is most sensitive) are rather small (see \reffig{fig:boost_several_prescriptions}).\ 

For comparison, we also show in \reffig{fig:constraints_est_vs_bbks} the $\gamma$-ray constraints from D18, using both point-sources and the diffuse flux (this corresponds to the gray solid line in \reffig{fig:compilation_constraints}). We note that these constraints were derived with the BBKS formalism (considering only the spiky component) and assuming the same DM mass, annihilation channel and cross-section as in our analysis. Hence, comparing the black and red dashed curves in  \reffig{fig:constraints_est_vs_bbks} is an useful exercise in order to estimate which observable (CMB anisotropy or $\gamma$-rays) has more constraining power on the small-scale primordial spectrum.  We see that the CMB constraints are weaker for $\ks < 10^4 \ \rm{Mpc}^{-1}$ (where the search for point-sources in $\gamma$-rays dominates the constraints) but they become competitive with those from the $\gamma$-ray diffuse flux for larger values of $\ks$. Nevertheless, this comparison should be taken with caution, as we have argued that any realistic computation of the boost factor should include the effect of late-time halo mergers and thus consider both the smooth power-law $\Lambda$CDM and spiky components in the primordial spectrum \footnote{Even though these cautions can be mitigated if UCMHs survive late-time accretion and mergers (in part as subhalos producing a possible subhalo boost to the total boost), this is still expected to have an impact on low-$z$ astrophysical observations, like the 21-cm signal but also $\gamma$-rays. In order to estimate a conservative constraining power of $\gamma$-rays, one could conduct the same analysis as in D18 but evaluating the ultracompact subhalo population and properties in the Milky Way with merger tree algorithms based on our improved EST formalism described in \refsec{sec:EST_HMF} (see Ref.~\cite{Ando:2022tpj} as example of an analysis adopting a similar approach).}. Let us add that our constraints are broadly in good agreement (although slightly stronger) than those of Ref.~\cite{Kawasaki:2021yek}, which also considered the impact of $s$-wave annihilations on the CMB for the same DM mass and cross-section as in our study. We attribute the residual differences to the distinct annihilation channel considered ($e^{+} e^{-}$ instead of $b \bar{b}$), as well as the distinct codes used ({\tt CAMB}/{\tt RECFAST} instead of {\tt ExoCLASS}/{\tt HYREC}). Let us also remark that Ref.~\cite{Kawasaki:2021yek} have found almost identical constraints between BBKS/EST (as opposed to our study), but this is of course because the authors considered neither the smooth $\Lambda$CDM component in $\mathcal{P}_{\mathcal{R}}(k)$ nor the combination of different density profiles when dealing with the EST approach.

\subsubsection{Considering other DM masses and annihilation channels}

\begin{figure}[t!]
\centering
\includegraphics[width=0.93\linewidth]{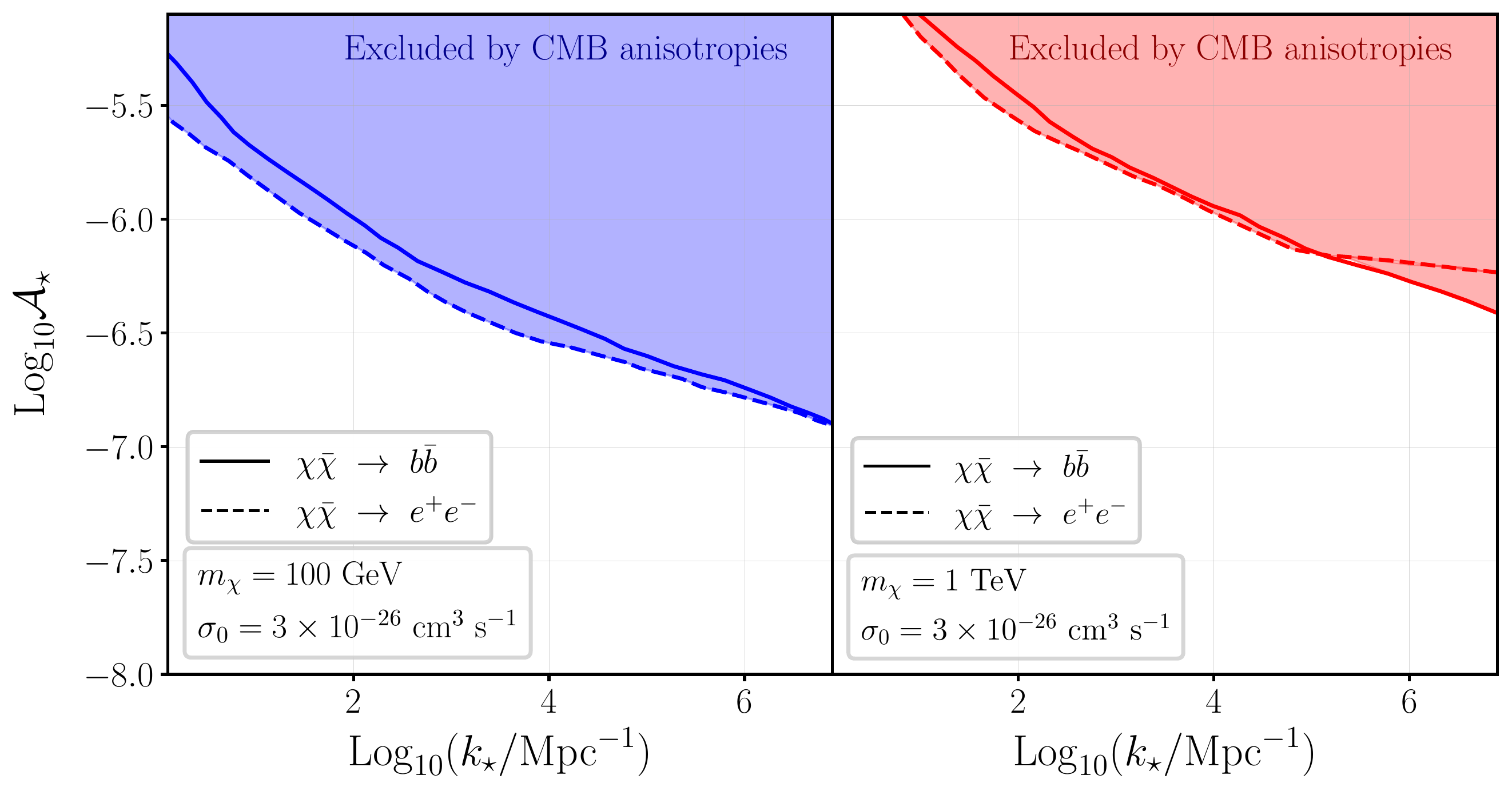} 
\caption{95\% C.L. bounds from \Planck+\BAO+\Pantheon~on the amplitude of the spike in the primordial spectrum as a function of its wavenumber, for different DM masses ($m_{\chi} =100 \ \rm{GeV}, \  1 \ \rm{TeV}$) and annihilation channels ($b \bar{b}$ and $e^{+} e^{-}$). These curves were obtained using the EST formalism and fixing the cross-section to the $s$-wave thermal relic value. }
\label{fig:constraints_several_m}
\end{figure}

As discussed in \refsec{sec:EnergyInjection}, the DM annihilation signal within halos depends not only on the effective boost factor $\mathcal{B}_\ell(z)$, but also on the annihilation parameter $p_{\rm ann,\ell} \equiv \sigma_\ell/m_\chi$ as well as the energy deposition functions $f_{\ell,c}(z)$ (which are ultimately determined by the annihilation channel as well as the DM mass $m_\chi$). Until now we have shown constraints on the primordial spectrum assuming a certain annihilation channel and fixed values of $\sigma_\ell$ and $m_\chi$, but it is interesting to vary these parameters in order to see how this could impact the constraints. In \reffig{fig:constraints_several_m}, we show the \Planck+\BAO+\Pantheon~bounds on $\As$ vs. $\ks$, for two different DM masses ($m_{\chi} =100 \ \rm{GeV}, \  1 \ \rm{TeV}$) and two different annihilation channels ($\chi \bar{\chi} \rightarrow b \bar{b}$ and $\chi \bar{\chi} \rightarrow  e^{+} e^{-}$). We have chosen these DM models simply because they serve to approximately bracket the uncertainty in the DM annihilation signal and also because they are allowed by standard \Planck~analyses \cite{Planck:2018vyg}. We see that the largest impact comes from the choice of DM mass: reducing $m_\chi$ by a factor of 10 tightens the constraints on $\As$ by roughly a factor of 3. On the contrary, the choice of DM annihilation channel has a very minor impact, which was already known for DM annihilations in the smooth background \cite{Planck:2018vyg}.

\subsection{The $p$-wave case }
\label{sec:results_pwave}

\begin{figure}[t!]
\centering
\includegraphics[width=0.7\linewidth]{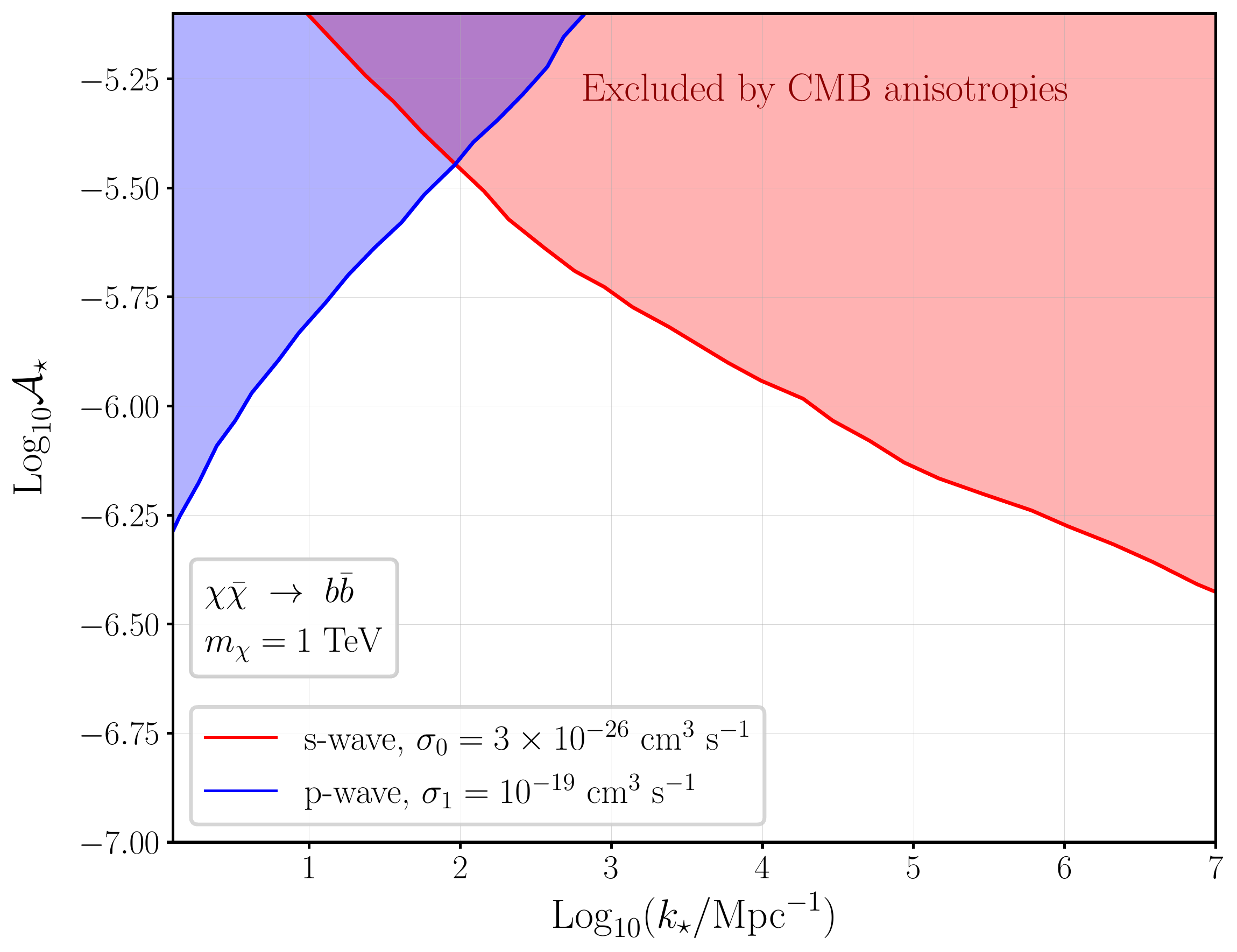} 
\caption{95\% C.L. bounds from \Planck+\BAO+\Pantheon~on the amplitude of the spike in the primordial spectrum as a function of its wavenumber, considering both $s-$ and $p-$wave DM annihilations. These curves were obtained using the EST formalism and assuming that DM annihilates into $b\bar{b}$ with a mass $m_{\chi} =1 \ \rm{TeV}$ and a cross-section $\sigma_0 = 3\times 10^{-26}~\rm cm^3 s^{-1}$ ($\sigma_1 = 10^{-19}~\rm cm^3s^{-1}$) for the $s-$wave ($p-$wave) case.}
\label{fig:constraints_sp_wave}
\end{figure}

Let us now move on to discuss the case of $p$-wave DM annihilations, which has always been ignored in past UCMH studies. Indeed, given the very small velocity of the smooth DM component, it is not obvious that this type of annihilations can yield any constraints on the primordial spectrum. However, the virialisation of early collapsed structures also increases the DM velocity with respect to the background, which could potentially have a non-negligible impact on the CMB \cite{Diamanti:2013bia}. In order to check this expectation, we computed for the first time the boost factor from $p$-wave annihilations in the presence of a spike in the primordial spectrum (see right panel in \reffig{fig:boost_several_prescriptions}). While we found that the $p$-wave boost has some support at $z \sim 10^3$ for $k_\star < 10^2 \ \mathrm{Mpc}^{-1}$, it is still bellow the $s$-wave boost by at least four orders of magnitude. Hence, picking the thermal value for the $p$-wave annihilation cross-section ($\sigma_1 \simeq 6 \times 10^{-26} \ \rm{cm}^3 \rm{s}^{-1}$) is not expected to produce any measurable effect. Nevertheless, we want to remain agnostic about the particle DM model and just check whether $p$-wave annihilations can produce any constraints at all on $\As$. Thus, we can allow ourselves to pick much larger values of $\sigma_1$, taking advantage of the fact that current constraints on $\sigma_1$ are significantly above the thermal relic value. In particular, we consider a model where DM annihilates into $b\bar{b}$ with a mass $m_{\chi} =1 \ \rm{TeV}$ and $\sigma_1 = 10^{-19}~\rm cm^3s^{-1}$, which is still one order of magnitude below current strongest astrophysical constraints \cite{Boudaud:2018oya}. \

In \reffig{fig:constraints_sp_wave} we show the corresponding \Planck+\BAO+\Pantheon~bounds on $\As$ vs. $\ks$, together with the $s$-wave bounds previously shown in \reffig{fig:constraints_est_vs_bbks}. We see that the constraints in the $p$-wave case are more prominent at small $\ks$, and gradually relax for larger $\ks$ values, which is exactly the opposite of what happens in the $s$-wave case. This behaviour was expected from the scalings of the 1-halo boosts $\mathcal{B}_{\ell, h}$ with $\ks$ that we discussed at the end of \refsec{sec:one_halo_boost}. Importantly, we find that $p$-wave annihilations can yield competitive constraints on the primordial spectrum at small $\ks$, which is particularly relevant for particle DM models that predict $\sigma_0 = 0$. We consider that this point is far from trivial and it is (to our knowledge) the first time that has been highlighted in the literature.

\section{Conclusion}
\label{sec:conclusion}
UCMHs constitute a very promising avenue to probe the primordial power spectrum $\mathcal{P}_{\mathcal{R}}(k)$ at scales $k \sim 1-10^7 \ \rm{Mpc}^{-1}$, where currently only upper limits exist. Because UCMH formation requires smaller enhancements in the primordial density field than those needed for PBH formation, the resulting constraints in $\mathcal{P}_{\mathcal{R}}(k)$ are much stronger, hence offering a valuable window into the early universe \cite{Aslanyan:2015hmi}. One effective way of testing the presence of UCMHs is to consider the effects of self-annihilating DM particles within these halos. More precisely, the earlier formation time and the compactness of UCMHs can significantly boost the DM annihilation signal and lead to a strong impact on different observables, including the $\gamma$-ray fluxes as well as the CMB anisotropies.\ 

In this work, we have considered a monochromatic enhancement in the primordial power spectrum (i.e., a ``spike"), which triggers the collapse of UCMHs at very large redshifts. We have performed
a thorough semi-analytical calculation of the cosmological boost factor based on the halo model, and quantified the impact of the boosted DM annihilation signal on the CMB anisotropies. Using our modified version of {\tt ExoCLASS} and data from \Planck+\BAO+\Pantheon, we have updated the CMB bounds on the amplitude $\As$ and wavenumber $\ks$ of the spike in the primordial spectrum. We have additionally provided a comprehensive compilation of the most stringent limits on $\mathcal{P}_{\mathcal{R}}(k)$ over a wide range of scales, as illustrated in \reffig{fig:compilation_constraints}. Our main findings can be summarized as follows: 

\begin{itemize}
    \item We have characterized the halo population using two distinct formalisms: BBKS (applied just to the spiky part of $\mathcal{P}_{\mathcal{R}}(k)$, as in past studies) and EST (applied for the first time to both the spiky and smooth $\Lambda$CDM parts of $\mathcal{P}_{\mathcal{R}}(k)$). While BBKS should provide a good description of UCMHs if they are isolated and have a negligible merger rate, we have argued that realistically one should consider a mixture made of UCMHs with Moore profiles and standard halos with NFW profiles, as a result of successive mergers. We have proved that this can be modeled using EST, and showed that the corresponding boost factors are generically smaller than in BBKS and hence lead to slightly weaker constraints. We have also used EST to estimate the theoretical uncertainty associated to the formation redshift. We have determined that the differences in the boost factor between BBKS and EST are always small at $z \sim 10^2 - 10^3$ (explaining why the impact on CMB constraints is minor) but can become much more significant at $z \lesssim 10^2$. This will be very relevant for testing the annihilation signal in UCMHs with low-$z$ astrophysical probes, like the 21-cm signal. 
    
    \item Assuming a fiducial model where DM annihilates into $b\bar{b}$ with a mass $m_\chi = 1 \ \rm{TeV}$ and a $s$-wave thermal cross-section, we have derived new CMB constraints on the amplitude and location of the spike in $\mathcal{P}_{\mathcal{R}}(k)$. Comparing with other UCMH studies, we see that $\gamma$-rays still dominate the constraints on $\mathcal{P}_{\mathcal{R}}(k)$ at $k \sim 10 -10^4 \ \rm{Mpc}^{-1}$, but our CMB constraints become very competitive at $k \gtrsim 10^4 \ \rm{Mpc}^{-1}$, ruling out primordial amplitudes $\As \gtrsim 10^{-6.5}$. In fact, after $\gamma$-rays, our constraints on $\mathcal{P}_{\mathcal{R}}(k)$ are the strongest available in the literature in the range $k \sim 10^5-10^6 \ \rm{Mpc}^{-1}$, which is valuable given that the CMB is very robust to astrophysical systematics. In order to bracket the uncertainty associated to the DM annihilation signal, we have also compared the constraints assuming two different annihilation channels ($\chi \bar{\chi} \rightarrow b \bar{b}, e^{+} e^{-}$) and two different DM masses ($m_{\chi} =100 \ \rm{GeV}, \  1 \ \rm{TeV}$). We have verified that the choice of the annihilation channel on the constraints is marginal, whereas the choice of DM mass can have an significant impact (namely, reducing the DM mass by a factor of 10 tightens the constraints on $\As$ by a factor of 3).

    \item We have computed the boost factor for $p$-wave DM annihilations in presence of the spike in $\mathcal{P}_{\mathcal{R}}(k)$. This case was ignored in past UCMH studies, since the very small velocity of the smooth DM component leads to a very suppressed DM annihilation signal. Nevertheless, the virialisation of early formed halos increases the DM velocity with respect to the background, which can enhance the DM annihilation signal at CMB times, as confirmed by our $p$-wave boost calculation. We have derived the first $p$-wave constraints on $\mathcal{P}_{\mathcal{R}}(k)$ and found that they are competitive at $k \sim 10 - 100 \ \rm{Mpc}^{-1}$ for DM models still allowed by current strongest astrophysical constraints (in particular, we considered a fiducial model with $\chi \bar{\chi} \rightarrow b \bar{b}$, $m_\chi = 1 \ \rm{TeV}$ and $\sigma_1 = 10^{-19} \rm{cm}^3 \rm{s}^{-1}$). The fact that $p$-wave has some constraining power on the primordial spectrum is very relevant for DM models that predict vanishing $s$-wave terms.
\end{itemize}
Our work highlights the synergy between UCMHs and DM annihilations, and the importance of UCMHs for probing the primordial power spectrum on small scales. In light of the new analytical EST-based treatment that we have developed here, it will be interesting to forecast the constraining power of future 21cm surveys on $\mathcal{P}_{\mathcal{R}}(k)$, as well as reexamine the $\gamma$-ray constraints by \cite{Delos:2018ueo}. Another interesting follow-up to this study will be to see the gravitational lensing signatures of UCMHs \cite{Delos:2023fpm}, as the derived constraints on the primordial spectrum would be independent of the WIMP hypothesis for DM. 

\acknowledgments
We would like to warmly thank Julien Lavalle and Vivian Poulin for their constructive comments, as well as their continued support and advice since this project started during a stay at LUPM (CNRS-IN2P3, Université de Montpellier). We acknowledge support from the GaDaMa ANR project (ANR-18-CE31-0006). GFA is supported by the European Research Council (ERC) under the European Union's Horizon 2020 research and innovation programme (Grant agreement No. 864035 - Undark). GF thanks Théo Moret for useful discussions on the halo distribution. GF acknowledges support of the ARC program of the Federation Wallonie-Bruxelles and of the Excellence of Science (EoS) project No. 30820817 - be.h “The H boson gateway to physics beyond the Standard Model”.


\appendix

\section{Injected energy and boost-factor for velocity dependent cross-sections}
\label{app:injvel}

In this appendix we detail the formulas related to energy injection and deposition from dark matter annihilating with a velocity-dependent cross-section into SM species.

\subsection{Energy injection from isotropic velocity distributions}
\label{app:injvel_dEdt}

We can generalise the partial wave expansion of the cross-section by writing
\begin{equation}
    \sigma v_{\rm rel} = \sum_{\ell = 0}^{\infty} \sigma_\ell v_{\rm rel}^{2\ell}
\end{equation}
The total injected energy per units of time and volume, at position $\bf r$ in the Universe, is thus
\begin{equation}
    \left.  \frac{{\rm d} E}{{\rm d} t {\rm d} V} \right|_{\rm inj.} ({\bf r}) = \sum_{\ell = 0}^{\infty}  \left.  \frac{{\rm d} E_\ell}{{\rm d} t {\rm d} V} \right|_{\rm inj.} ({\bf r})
\end{equation}
with the definition of the injected energy associated to a partial wave,
\begin{equation}
\begin{split}
   \left.  \frac{{\rm d} E_\ell}{{\rm d} t {\rm d} V} \right|_{\rm inj.} ({\bf r}) & =  \frac{\sigma_\ell}{m_\chi} \int {\rm d}^3 {\bf v}_1  {\rm d}^3 {\bf v}_2 f({\bf v}_1, {\bf r}) f({\bf v}_2, {\bf r}) ({\bf v}_2 - {\bf v_1})^{2\ell}  \, .
   \end{split}
\end{equation}
Here we denote the one point phase space distribution function (PSDF) of the dark matter by $f$. Assuming an isotropic the velocity distribution, $f({\bf v}, {\bf r}) = f(v, {\bf r})$ with $v = |{\bf v}|$, we can massage the expression of the injected energy under the form
\begin{equation}
\begin{split}
   \left.  \frac{{\rm d} E_\ell}{{\rm d} t {\rm d} V} \right|_{\rm inj.} ({\bf r}) & =  \frac{4\pi \sigma_\ell}{m_\chi} \int {\rm d}^3 {\bf v}_1  {\rm d} v_2 v_2^2 f(v_1, {\bf r}) f(v_2, {\bf r}) G(v_1, v_2)\, .
   \end{split}
\end{equation}
with the function
\begin{equation}
\begin{split}
    G(v_1, v_2) \equiv  & \frac{1}{2}\sum_{q=0}^\ell \binom{\ell}{q} (v_1^2+v_2^2)^{\ell-q}(2v_1v_2)^q \int_0^{\pi} {\rm d}\theta \sin \theta \cos^q \theta\\
        = & \frac{1}{2} \sum_{q=0}^\ell \binom{\ell}{q} (v_1^2+v_2^2)^{\ell-q}(2v_1v_2)^q \frac{1 + (-1)^q}{1+q} \, .\\
\end{split}
\end{equation}
Massaging the binomial expression it yields
\begin{equation}
    \begin{split}
    G(v_1, v_2) = &  \frac{1}{2(\ell+1)}\sum_{q=0}^\ell \binom{\ell+1}{q+1} (v_1^2+v_2^2)^{\ell-q}(2v_1v_2)^q \left[1 + (-1)^q\right]\\
        = &  \frac{1}{2(\ell+1)} \sum_{p=0}^{\ell} \binom{\ell+1}{p} (v_1^2+v_2^2)^{\ell+1-p}(2v_1v_2)^{p-1} \left[1 - (-1)^{p}\right] \\
        = & \frac{1}{2v_1v_2} \frac{1}{2(\ell + 1)}\left[(v_1 + v_2)^{2(\ell+1)} - (v_1 - v_2)^{2(\ell+1)}\right]
    \end{split}
\end{equation}
Further expanding in series and changing variables gives
\begin{equation}
\begin{split}
    G(v_1, v_2) = & \frac{1}{2v_1v_2} \frac{1}{2(\ell + 1)} \sum_{q=0}^{2(\ell+1)} \binom{2(\ell+1)}{q} v_1^q v_2^{2(\ell+1)-q} \left[1-(-1)^q\right]\\
    G(v_1, v_2) = & \frac{1}{2(\ell + 1)} \sum_{q=0}^{\ell} \binom{2(\ell+1)}{2k+1} v_1^{2k} v_2^{2(\ell-k)} \, .
    \end{split}
\end{equation}
Putting everything back together, one has
\begin{equation}
      \boxed{\left.  \frac{{\rm d} E_\ell}{{\rm d} t {\rm d} V} \right|_{\rm inj.} ({\bf r}) =  \rho_\chi^2({\bf r}) \frac{\sigma_\ell}{2m_\chi}\frac{1}{\ell+1}\sum_{k=0}^{\ell} \binom{2(\ell+1)}{2k+1} \left< v^{2(\ell-k) } \right> \left< v^{2k} \right> }
\end{equation}
This expression depends on moments of the velocity as defined in \refequ{eq:velocity_moment}. 

\subsection{Definition of the boost factor}
\label{app:injvel_boost}

In the main text, in \refequ{eq:Def_Rl}, we have introduced the ratio of the true injected energy over the injected energy in a scenario where all the DM is smooth and annihilates with an $s$-wave cross-section. This ratio depends on the space-averaged injected energy. To properly define what this means, let us consider a function $g : {\bf r} \to g ({\bf r})$. The space-averaged value of $g$ is defined by
\begin{equation}
    \overline{g} \equiv \lim_{V\to \infty} \frac{1}{V}\int_{V} g ({\bf r}) {\rm d}^3 {\bf r}\, .
    \label{eq:spaceAverage}
\end{equation}
Thus, $\mathcal{R}_\ell$ takes the general form
\begin{equation}
    \mathcal{R}_\ell(z) = \frac{1}{2(\ell+1)}\sum_{k=0}^{\ell}\binom{2(\ell+1)}{2k+1} R_{\ell, k}(z)\, ,
\end{equation}
with each term in the sum being
\begin{equation}
    R_{\ell, k}(z) \equiv  \frac{1}{\overline{\rho_\chi}^2(z)}\int \overline{f({\bf v}_1, {\bf r}) f({\bf v}_2, {\bf r})} v_1^{2(\ell-k)}v_2^{2k} {\rm d}^3 {\bf v}_1 {\rm d}^3 {\bf v}_2 \, .
\end{equation}
The total PSDF can be written as a sum of two components (one from the smooth homogeneous background and one from the halos). Nonetheless, they are not both isotropic in the same frame due to the relative different of halos over the smooth background. In the following we consider $f_{\rm sm}(v)$ for the smooth part -- where the velocity is defined with respect to the cosmic frame, and $f_{\rm h}(v, {\bf r})$ for halos where the velocity within each halo is defined with respect to its proper motion. From these we define 
\begin{equation}
  \forall n \in \mathbb{N} \quad 
\begin{cases}
    \displaystyle  \left< v^n \right>_{\rm sm} \equiv \frac{1}{\rho_{\rm sm}} \int f_{\rm sm}(v) v^n {\rm d}^3 {\bf v} \quad  {\rm with} \quad \rho_{\rm sm} = \int f_{\rm sm}(v) {\rm d}^3 {\bf v}\\[10pt]
     \displaystyle  \left< v^n \right>_{\rm h} \equiv \frac{1}{\rho_{\rm h}({\bf r})} \int f_{\rm h}(v, {\bf r}) v^n {\rm d}^3 {\bf v} \quad  {\rm with} \quad \rho_{\rm h}({\bf r}) = \int f_{\rm h}(v, {\bf r}) {\rm d}^3 {\bf v}\, .
    \end{cases}
\end{equation}
and we can introduce the boost as
\begin{equation}
    B_{\ell, k}(z) \equiv  \frac{1}{\overline{\rho_\chi}^2(z)}  \overline{\rho_{\rm h}^2 \left< v^{2k}\right>_{\rm h}\left< v^{2(\ell-k)}\right>_{\rm h} } \, .
\end{equation}
The effective boost defined in the main text is therefore
\begin{equation}
        \mathcal{B}_\ell(z) = \frac{1}{2(\ell+1)}\sum_{k=0}^{\ell}\binom{2(\ell+1)}{2k+1} B_{\ell, k}(z) \, .
\end{equation}
Because of the product of PSDFs in the integrals (or the product of velocity moments) one cannot properly evaluate the cross term, corresponding to the annihilation of a particle from an halo with a particle from the background. Such cross-terms are usually omitted as we have done in the main text. For the sake of the exercise, let us consider here that halos are at rest (or with velocity comparable to that of the smooth background much smaller than the velocity of the particles inside due to virtualisation). Then we can consider that the center of mass frame of the halos is similar to the cosmic rest frame and we can write
\begin{equation}
    f(v, {\bf r}) \sim f_{\rm sm}(v) + f_{\rm h}(v, {\bf r}) \, .
\end{equation}
Therefore one can massage the expression above as
\begin{equation}
\begin{split}
    R_{\ell, k}(z)  = & B_{\ell, k}(z) +  \frac{\rho_{\rm sm}^2(z)}{\overline{\rho_\chi}^2(z)} \left< v^{2(\ell-k)}\right>_{\rm sm} \left< v^{2k}\right>_{\rm sm}  \\
    & + \frac{\rho_{\rm sm}(z)}{\overline{\rho_\chi}^2(z)}  \left< v^{2(\ell-k)}\right>_{\rm sm} \overline{\rho_{\rm h} \left< v^{2k}\right>_{\rm h} } \\
    & + \frac{\rho_{\rm sm}(z)}{\overline{\rho_\chi}^2(z)}   \left< v^{2k}\right>_{\rm sm} \overline{\rho_{\rm h} \left< v^{2(\ell-k)}\right>_{\rm h} }
\end{split}
\end{equation}
For all quantities that are halo dependent, the space-averaged value can be obtained from the halo model. Similarly to \refequ{eq:spaceAverage}, let us consider a function $g_{\rm h}$ that represents a quantity associated to halos. The space-averaged value of $g_{\rm h}$ can be aproximated by
\begin{equation}
      \overline{g_{\rm h}} \equiv \lim_{V\to \infty} \frac{1}{V}\int_{V} g_{\rm h} ({\bf r}) {\rm d}^3 {\bf r} \simeq (1+z)^3 \int {\rm d} M {\rm d} z_{\rm f} \frac{\partial^2 n(z)}{\partial M \partial z_{\rm f}} \int_0^{r_\Delta(z)} g_{\rm h}({\bf r}) {\rm d}^3 {\bf r}
\end{equation}
Thus, one has
\begin{equation}
\boxed{\begin{split}
   & \overline{\rho_{\rm h} \left< v^n\right>_{\rm h} } = \int {\rm d} M {\rm d} z_{\rm f} \frac{\partial^2 n(z)}{\partial M \partial z_{\rm f}} \int_0^{r_\Delta(z)} \rho_{\rm h}({\bf r}) \left< v^n\right>_{\rm h} {\rm d}^3 {\bf r} \\
   & B_{\ell, k}(z) = \frac{(1+z)^3}{\overline{\rho_\chi}^2(z)}\int {\rm d} M {\rm d} z_{\rm f} \frac{\partial^2 n(z)}{\partial M \partial z_{\rm f}} \int_0^{r_\Delta(z)} \rho^2_{\rm h}({\bf r}) \left< v^{2k}\right>_{\rm h} \left< v^{2(\ell-k)}\right>_{\rm h} {\rm d}^3 {\bf r} 
   \end{split}}
\end{equation}

\subsection{The $s$- and $p$-wave cases}
\label{app:injvel_sp}

We have derived general formulas above, let us apply them to the case of the $s$- and $p$-wave scenario that are the focus of this work. We consider both cases below.\\

\noindent
{\bf The $s$-wave case}: \\
For $\ell = 0$, one has $\mathcal{R}_0 = R_{0, 0}$ and $\mathcal{B}_0 = B_{0, 0}$. Therefore,
\begin{equation}
    \mathcal{R}_0(z) = \mathcal{B}_0(z) + \frac{\rho_{\rm sm}^2(z)}{\overline{\rho_\chi}^2(z)} + 2\frac{\rho_{\rm sm}(z)\overline{\rho_{\rm h}}(z)}{\overline{\rho_\chi}^2(z)} 
\end{equation}
where, by conservation of matter, the average density in halos must satisfy $\overline{\rho_{\rm h}} = \overline{\rho_{\chi}} - \rho_{\rm sm}$. Therefore we can simplify this expression to
\begin{equation}
\boxed{
    \mathcal{R}_0(z) = 1 - \left(1-\frac{\rho_{\rm sm}(z)}{\overline{\rho_{\chi}}(z)}\right)^2 + \mathcal{B}_{0}(z)} \, .
\end{equation}
The boost factor is
\begin{equation}
    \mathcal{B}_0(z) = \frac{(1+z)^3}{\overline{\rho_\chi}^2(z)} \int {\rm d} M {\rm d} z_{\rm f} \frac{\partial^2 n(z)}{\partial M \partial z_{\rm f}} \int_0^{r_\Delta(z)} \rho^2_{\rm h}({\bf r}) {\rm d}^3 {\bf r} \, .
\end{equation}
Note that in this scenario, because the cross-section does not depend on the velocity, the expressions above are actually exact whether or not the assumption on the halo velocity holds or not. \\

\noindent
{\bf The $p$-wave case}: \\
For $\ell = 1$, one has $\mathcal{R}_1 = R_{1, 0} + R_{1,1} = 2R_{1,0}$ and $\mathcal{B}_1 = B_{1, 0} + B_{1,1} = 2B_{1,0}$. Therefore,
\begin{equation}
    \mathcal{R}_1(z) = \mathcal{B}_1(z) + 2\frac{\rho_{\rm sm}^2(z)}{\overline{\rho_\chi}^2(z)}\left<v^2\right>_{\rm sm} + 2\frac{\rho_{\rm sm}(z)\overline{\rho_{\rm h}}(z)}{\overline{\rho_\chi}^2(z)}  \left< v^{2}\right>_{\rm sm}  + 2\frac{\rho_{\rm sm}(z)}{\overline{\rho_\chi}^2(z)}  \overline{\rho_{\rm h} \left< v^{2}\right>_{\rm h} }
\end{equation}
Using that $\overline{\rho_{\rm h}} = \overline{\rho_{\chi}} - \rho_{\rm sm}$ and defining the average velocity dispersion in halos as,
\begin{equation}
    \overline{\left< v^2 \right>_{\rm h}} \equiv  \frac{1}{\overline{\rho_{\rm h}}}\overline{\rho_{\rm h} \left< v^{2}\right>_{\rm h} }\, ,
\end{equation}
it yields the simplified expression,
\begin{equation}
\boxed{
    \mathcal{R}_1(z) \simeq 2\left(\frac{\rho_{\rm sm}(z)}{\overline{\rho_\chi}(z)}\right)^2\left<v^2\right>_{\rm sm}  +  2\frac{\rho_{\rm sm}(z)}{\overline{\rho_\chi}(z)} \left\{1-\frac{\rho_{\rm sm}(z)}{\overline{\rho_\chi}(z)}\right\}\left[\left<v^2\right>_{\rm sm} + \overline{\left<v^2\right>}_{\rm h}\right] + \mathcal{B}_{1}(z) \, .}
\end{equation}
The boost factor is
\begin{equation}
    \mathcal{B}_1(z) = \frac{(1+z)^3}{\overline{\rho_\chi}^2(z)} \int {\rm d} M {\rm d} z_{\rm f} \frac{\partial^2 n(z)}{\partial M \partial z_{\rm f}} \int_0^{r_\Delta(z)} \rho^2_{\rm h}({\bf r}) \left[2 \left< v^2\right>_{\rm h} \right] {\rm d}^3 {\bf r} \, .
\end{equation}\

Eventually, halo collapse and formation is always viewed considering the total mass density (as the entire mass contributed to the formation of structure). To that reason the mass function in the EST approach is normalised to the total mass density of the Universe and runs over the total mass of the structures. Integrating over the halo profiles one recover that cosmological mass. Therefore to properly define the DM density, from the virial mass and concentration one should rescale the density profile by
\begin{equation}
    \rho_{\rm h}(r, z) \longrightarrow \frac{\Omega_{\chi, 0}}{\Omega_{\rm m, 0}} \rho_{\rm h}(r, z) \, .
\end{equation}
In our case, we equivalently rescale the expression of the boost factor -- the only quantity that matters in this analysis, as shown by \refequ{eq:Rlsimplsimpl}. One get the same result as if we had rescaled the density by changing the denominator from  $\overline{\rho_{\chi}}^2(z)$ to $\overline{\rho_{\rm m}}^2(z)$ 
\begin{equation}
\boxed{
    \mathcal{B_{\ell}}(z) \longrightarrow \frac{(1+z)^3}{\overline{\rho_{\rm m}}^2(z)} \int {\rm d} M {\rm d} z_{\rm f} \frac{\partial^2 n(z)}{\partial M \partial z_{\rm f}} \int_0^{r_\Delta(z)} \rho^2_{\rm h}({\bf r}) \left[2 \left< v^2\right>_{\rm h} \right]^\ell {\rm d}^3 {\bf r} \, . }
\end{equation}
Note that the rescaling does not impact of the velocity distribution as the latter depends on the total mass content in the structure. Let us now detail the expressions of the luminosity functions.

\subsection{Dimensionless mass and luminosity}

According to \refequ{eq:dimensionlessOneHaloBoost}, integrals entering in the expression of the 1-halo boost factor are:
\begin{equation}
 \mu(x) \equiv \int_{0}^{x} \tilde \rho(y) y^2 {\rm d}y, \ \ \ \ \ \ \lambda_0(x) \equiv \int_{0}^{x} \tilde \rho^2(y) y^2 {\rm d}y,
\end{equation}
\begin{equation}
    \lambda_{1}(x) \equiv  \int_0^{x} {\rm d} y y^2  \tilde \rho(y) \int_y^{c} {\rm d} y'  \frac{\tilde \rho(y')\mu(y')}{y'^2}  \, ,
\end{equation}
where $\tilde \rho(y) \equiv y^{-\gamma}(1+y)^{-3+\gamma}$. To simplify the latter expression when evaluated for $x=c$ (as it is always the case in this analysis since we assume that halos extend up to their virial radius), by switching the order of the two integrals one finds
\begin{equation}
 \lambda_1(c) \equiv \int_{0}^{c} \frac{ \tilde\rho(y) \mu^2(y)}{y^2} {\rm d}y \, .
\end{equation}
Let us particularize for each density profile.\\

\noindent
{\bf NFW profile ($\gamma =1$)}: \
\begin{align}
\mu (c) &= \log{(1+c)} - \frac{c}{1+c}. \label{eq:mu_nfw}\\[0.8em]
\lambda_0 (c) &= \frac{c^3}{3} \left[ 1-\frac{1}{(1+c)^3}\right]. \label{eq:l0_nfw} \\[0.8em]
\lambda_1 (c) &= \log^2 (1+c) \left[ 3\log{ \left( \frac{1}{1+c}\right)} + \frac{2+3c}{c(1+c)} - \frac{1}{2} (1+c^{-2}) \right] \nonumber \\ 
 &-{\rm Li}_2 (-c) - 6 {\rm Li}_3 \left( \frac{1}{1+c} \right) -6 \log {(1+c)} {\rm Li}_2 \left( \frac{1}{1+c} \right) + 6 \zeta (3) - \frac{53}{6} \nonumber \\ 
&+ \log{(1+c)} \left[ \frac{1}{(1+c)^2} + \frac{1+7c}{c(1+c)} \right] + \frac{1}{1+c} \left[ 7 + \frac{1}{(1+c)} + \frac{1}{3(1+c)^2} \right]. \label{eq:l1_nfw}
\end{align}

\noindent
{\bf Moore profile ($\gamma =3/2$)}: \
\begin{align}
\mu (c) &= 2 \arcsinh{(\sqrt{c})} -2 \sqrt{\frac{c}{1+c}}. \label{eq:mu_moore} \\[0.8em]
\lambda_0 (c) &= \frac{1}{3} + \frac{2c+3}{2(1+c)^2} + \log{ \left(\frac{c}{1+c} \right)} - \frac{2D^{-1} +3}{2(1+D^{-1})^2} + \log{(1+D)}. \label{eq:l0_moore}  \\[0.8em] 
\lambda_1 (c) &= \frac{8}{15c^{5/2} (1+c)^{3/2}} \biggl[ c \left[-3+c\left( 9 + c (87 + 70c)\right)\right] - 7\pi^2 \sqrt{1+c} \left( c^{5/2} + c^{7/2} \right) \nonumber \\
&-6 \left(\sqrt{c^5(1+c)}+\sqrt{c^7(1+c)}\right) \left( 5 {\rm Li}_2 \left[ e^{-4\arcsinh{(\sqrt{c})}} \right] -12 {\rm Li}_2 \left[ e^{-2\arcsinh{(\sqrt{c})}} \right] \right)  \nonumber \\
&-3\arcsinh{(\sqrt{c})} \biggl[ -2\sqrt{c(1+c)}+5\sqrt{c^3(1+c)} + 12 \sqrt{c^5(1+c)}  \nonumber  \\ 
&+ \arcsinh{(\sqrt{c})} \left[ 1-16\sqrt{c^5(1+c)} -16 \sqrt{c^7(1+c)}   + c(1+2c) \left( -1 + 8c(1+c) \right) \right] \nonumber \\ 
& +8c^{5/2} (1+c)^{3/2} \left( \log{\left[1-e^{-2\arcsinh{(\sqrt{c})}} \right]} -5 \log{\left[1+e^{-2\arcsinh{(\sqrt{c})}} \right]} \right)
\biggr] \biggr]. \label{eq:l1_moore}
\end{align}
In \refequ{eq:l1_nfw} and \refequ{eq:l1_moore}, ${\rm Li}_2$ and ${\rm Li}_3$ denote the polylogarithms of order 2 and 3, while in \refequ{eq:l0_moore} we have introduced $D \equiv (\rho_{\rm max
}/\rho_{\rm s })^{2/3}$ to regularize the integral (see \refequ{eq:rho_maxx}).

\section{Mass function and formation redshift in the EST approach}
\label{app:HeavisideInPS}

In this appendix we first show a careful computation of the mass function from a non-continuous smoothed variance of the matter density field. Then, we check our results by computing the total mass density of the Universe from the expression we obtain. We end this section by extending the discussion on the formation redshift. 

\subsection{Regularisation of the Heaviside distribution}

Here we rely on tools from the theory of distributions and therefore use more formal notations when necessary. Using the sharp $k$-space filter, the smoothed variance associated to the mass $M$ can be written under the form of a step function -- see \refequ{eq:VarianceWithSpike},
\begin{equation}
    S_0 : M \mapsto  \alpha(M) + \beta \theta(M)
\end{equation}
where $\beta$ is a constant and $\alpha$ a smooth decreasing function of $M$ (of class $\mathcal{C}^\infty$). For compactness here we denote $\theta : M \mapsto \Theta(1 -M/M_\star)$ the Heaviside function centered on $M_\star$. In the PS formula $S_0$ appears together with its derivative which, thereby combined, define a non-trivial distribution. We introduce
\begin{equation}
    T : M \mapsto   \frac{\omega}{2 S_0^{3/2}(M)}\left| \frac{{\rm d}  S_0(M)}{{\rm d} M} \right| f\left(\frac{\omega}{S_0^{1/2}(M)} \right) \, .
    \label{eq:defTdist}
\end{equation}
Such a distribution can only be properly evaluated from its action on a test function $\varphi$. The first step is to define a regularised version of $T$, to write the action as the limit of an integral of smooth functions. The irregular bit from which difficulties arise comes from the Heaviside function. We introduce $h$, a $\mathcal{C}^{\infty}$ monotonic function satisfying
\begin{equation}
    \lim_{x \to -\infty} h(x) = 0 \quad {\rm and} \quad  \lim_{x \to +\infty} h(x) = 1 \, .
\end{equation}
An example is given by the logistic function. We then define the regularised Heaviside function building on the function $h$ by 
\begin{equation}
    \theta_\epsilon : M \mapsto  h\left(\frac{1}{\epsilon}\left\{1-\frac{M}{M_\star}\right\}\right) \quad \text{such that} \quad \theta = \lim_{\epsilon \to 0^+} \theta_\epsilon
\end{equation}
at least point-wise. Therefore we can define a family of smooth functions $T_\epsilon$ which tend to $T$ in the same limit. For now, let us fix $\epsilon > 0$. The distributions $T_\epsilon$ are defined by replacing the Heaviside function in the variance by its regularised counterpart,
\begin{equation}
\begin{split}
T_\epsilon \equiv  - \frac{\omega}{2}\frac{\alpha' + \beta \theta_\epsilon'}{[\alpha + \beta \theta_\epsilon]^{3/2}}  f\left(\frac{\omega}{[\alpha + \beta \theta_\epsilon]^{1/2}} \right)  \, .
\end{split}
\label{eq:Tepsilon}
\end{equation}
Note that the minus sign in front comes from the absolute value around the derivative of the variance, and from the fact that the derivative is negative. By definition, the action of the smooth function $T_\epsilon$ on the test function $\varphi$ is
\begin{equation}
\begin{split}
    \left< T_\epsilon , \varphi \right> & \equiv \int_0^{\infty} T_\epsilon\left(M\right) \varphi(M) {\rm d} M\, . 
\end{split}
\label{eq:actionTepsilon}
\end{equation}
and the action of $T$ is set from
\begin{equation}
    \left< T , \varphi \right> = \lim_{\epsilon \to 0^+}\left< T_\epsilon , \varphi \right> \, .
\end{equation}
We now introduce \refequ{eq:Tepsilon} into \refequ{eq:actionTepsilon} and decompose the expression into two parts, respectively corresponding to the derivative of the smooth part $\alpha$ and to the derivative of the regularised Heaviside $\theta_\epsilon$
\begin{equation}
    \begin{split}
    \left< T_\epsilon , \varphi \right> = \left< -\frac{\omega}{2} \frac{\alpha'}{[\alpha+\beta \theta_\epsilon]^{3/2}}f\left(\frac{\omega}{[\alpha + \beta \theta_\epsilon]^{1/2}}\right), \varphi \right >
        +  C_\epsilon \, ,
    \end{split}
    \label{eq:ActionTOnPhi}
\end{equation}
with the definition 
\begin{equation}
    C_\epsilon \equiv - \frac{\beta \omega}{2} \int_{0}^{\infty} \theta_\epsilon'(M)\frac{ f\left(\frac{\omega}{[\alpha(M) + \beta \theta_\epsilon(M)]^{1/2}} \right)}{[\alpha(M) + \beta \theta_\epsilon(M)]^{3/2}}\varphi(M) {\rm d} M \, .
\end{equation}
The first term in \refequ{eq:ActionTOnPhi} is the action of a regular function whose limit at $\epsilon \to 0^+$ is straightforwards. On the contrary, the limit of the second term cannot be evaluated directly as a divergence would appear in $\theta_\epsilon'$ at $M=M_\star$. Let us show that by re-writting the expression, the divergence vanishes. We introduce the change of variable $y : M \to (1 - M/M_\star)/\epsilon$. By definition of $\theta_\epsilon$ it yields
\begin{equation}
    C_\epsilon \equiv  \frac{\beta \omega}{2} \int_{-\infty}^{1/\epsilon} h'(y) \frac{ f\left(\frac{\omega}{[\alpha(M_\star(1-y \epsilon)) + \beta h(y)]^{1/2}} \right)}{[\alpha(M_\star(1 - y \epsilon)) + \beta h(y)]^{3/2}}\varphi(M_\star(1 - y \epsilon)) {\rm d} y \, .
\end{equation}
As $h'$ tends to 0 at $\pm \infty$, taking the limit $\epsilon$ small and performing successive changes of variable in the integrals then gives 
\begin{equation}
\begin{split}
\lim_{\epsilon \to 0^+} C_{\epsilon} & = \frac{\beta \omega}{2} \varphi(M_\star) \int_{-\infty}^{+\infty} h'(y) \frac{ f\left(\frac{\omega}{[\alpha(M_\star) + \beta h(y)]^{1/2}} \right)}{[\alpha(M_\star) + \beta h(y)]^{3/2}} {\rm d} y\\
& = \frac{\beta \omega}{2} \varphi(M_\star) \int_{0}^{1} \frac{f\left(\frac{\omega}{[\alpha(M_\star) + cw]^{1/2}} \right)}{[\alpha(M_\star) + \beta w]^{3/2}}  {\rm d} w \\
    & = \varphi(M_\star)  \int_{\omega/\sqrt{\beta+\alpha(M_\star)}}^{\omega/\sqrt{\alpha(M_\star)}} f(\nu) {\rm d} \nu \, .
    \end{split}
\end{equation}
The action of $T$ on $\varphi$ is thus
\begin{equation}
    \begin{split}
        \left< T, \varphi \right> & = \left< \frac{-\alpha'\omega f\left(\frac{\omega}{[\alpha + \beta \theta]^{1/2}}\right) }{2[\alpha+\beta \theta]^{3/2}}, \varphi  \right> + \lim_{\epsilon \to 0^+} C_\epsilon  \\
        & = \left< \frac{-\alpha' \omega f\left(\frac{\omega}{[\alpha + \beta \theta]^{1/2}}\right) }{2[\alpha+\beta \theta]^{3/2}} + \delta_{M_\star}   \int_{\omega/\sqrt{\beta+\alpha}}^{\omega/\sqrt{\alpha}} f(\nu) {\rm d} \nu, \varphi \right > \, ,
    \end{split}
\end{equation}
with the shorthand notation for the Dirac distribution $\delta_{M_\star} : M \to \delta(M-M_\star)$. In the end, the distribution $T$ takes the non trivial form
\begin{equation}
    T = \frac{-\alpha'\omega f\left(\frac{\omega}{[\alpha + \beta \theta]^{1/2}}\right) }{2[\alpha+\beta \theta]^{3/2}} + \delta_{M_\star}  \int_{\omega/\sqrt{\beta+\alpha}}^{\omega/\sqrt{\alpha}} f(\nu) {\rm d} \nu \, .
    \label{eq:Tdist}
\end{equation}

\subsection{Normalisation to the density of the Universe}

In this second subsection we show that the result of \refequ{eq:Tdist} satisfies to a normalisation condition required for mass functions. We also argue that the following proof is another way (less formal) to find the same result. Let us consider an idealized universe in which the variance satisfies
\begin{equation}
    \lim_{M \to -\infty} \alpha = +\infty \quad {\rm and} \quad \lim_{M\to 0} \alpha(M) = 0
\end{equation}
This implies that all matter should be in the form of halos that have collapse at some point in time before today (note that this assumption is not far from the reality of the hierarchical and fractal structure formation picture). Thus, in this universe the integral, due to the definition of $T$ in \refequ{eq:defTdist}, 
\begin{equation}
    \mathcal{N} \equiv \int_{0}^{+\infty} T(M) {\rm d} M 
\end{equation}
should satisfy $\mathcal{N}=1$. Let us show that this is indeed the case. First, by writing the expression for $T$ found above in the integral we get
\begin{equation}
   \mathcal{N} = -\int_{[0, M_\star[} \frac{\alpha'\omega f\left(\frac{\omega}{[\alpha + \beta]^{1/2}}\right) }{2[\alpha+\beta]^{3/2}} {\rm d} M   -\int_{]M_\star, \infty)} \frac{\alpha'\omega f\left(\frac{\omega}{\alpha^{1/2}}\right) }{2\alpha^{3/2}} {\rm d } M + \int_{\omega/\sqrt{\beta+\alpha(M_\star)}}^{\omega/\sqrt{\alpha(M_\star)}} f(\nu) {\rm d} \nu \, .
\end{equation}
This expression can be simplified by noticing that for any constant $c \ge 0$ one has
\begin{equation}
    \frac{{\rm d}}{{\rm d}M} \int^{\omega/\sqrt{c + \alpha}} f(\nu) {\rm d} \nu = - \frac{\alpha' \omega}{2 [\alpha + c]^{3/2}} f\left(\frac{\omega}{[\alpha +c]^{1/2}}\right)\, ,
\end{equation}
which yields the much simpler formula
\begin{equation}
    \mathcal{N} = \left[ \int^{\omega/\sqrt{\alpha + \beta}} f(\nu) {\rm d} \nu\right]_{0}^{M_\star} + \left[ \int^{\omega/\sqrt{\alpha}} f(\nu) {\rm d} \nu\right]_{M_\star}^{+\infty} + \int_{\omega/\sqrt{\beta+\alpha(M_\star)}}^{\omega/\sqrt{\alpha(M_\star)}} f(\nu) {\rm d} \nu  \, .
\end{equation}
Assuming the theoretical bounds on $\alpha$ given above, the expression simplifies to
\begin{equation}
    \mathcal{N} = \int_0^{+\infty} f(\nu) {\rm d} \nu \, .
\end{equation}
which readily gives $\mathcal{N}=1$ by construction of the EST/PS mass function. As mentioned in the beginning of this subsection, we could have inferred  \refequ{eq:Tdist} directly by following this argument upside-down. Due to the derivative of the Heaviside function in the definition of $T$ we know that it has to be the sum of a smooth function and a constant multiplying a Dirac distribution. Asking for the proper normalisation then automatically gives the value of this constant. We believe though that the more mathematical approach performed above establishes the result on more solid grounds.

\subsection{Discussion on the formation redshift}
\label{app:FormationRedshift}

\begin{figure}[t!]
\centering
\includegraphics[width=0.49\linewidth]{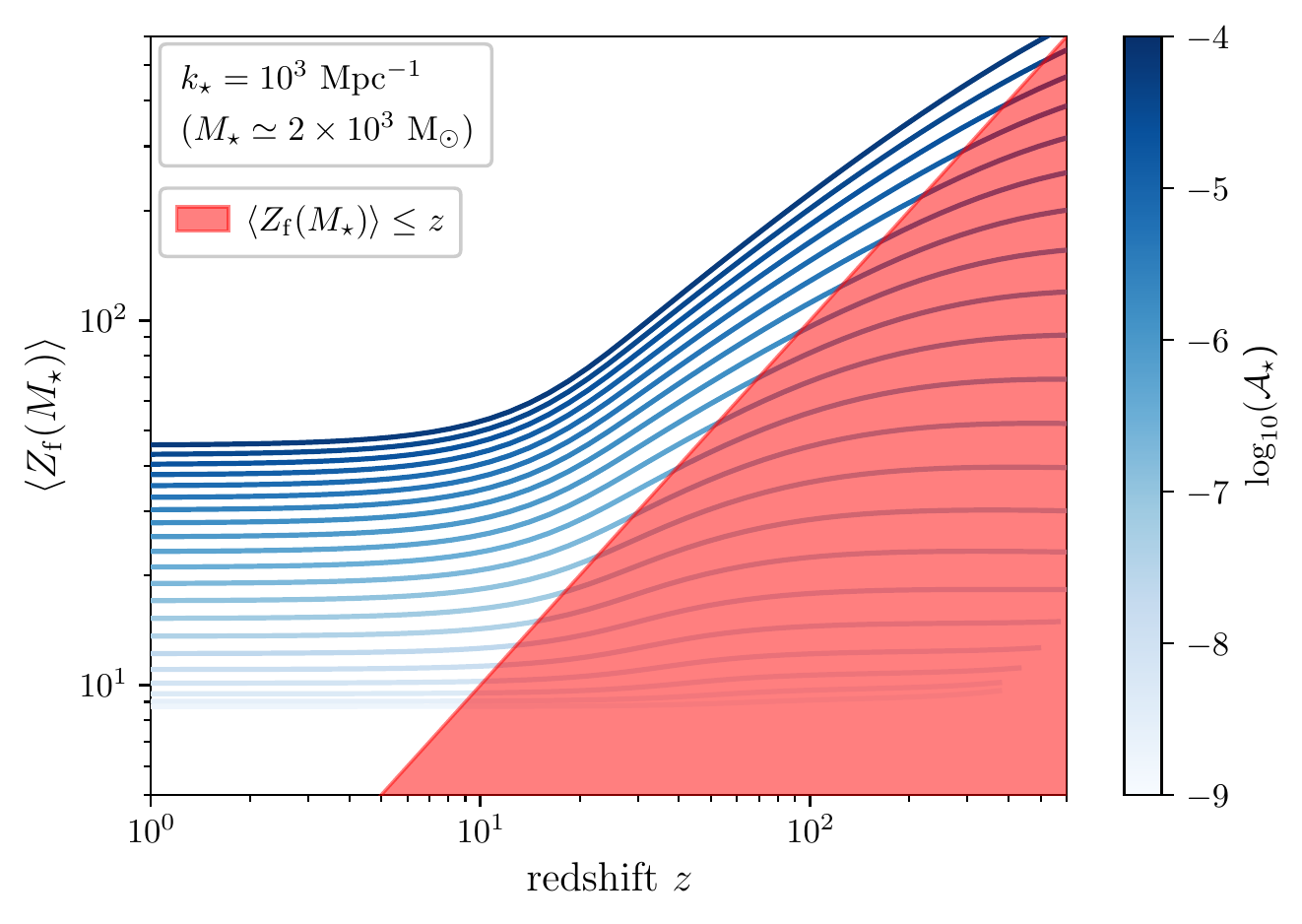} \hfill
\includegraphics[width=0.49\linewidth]{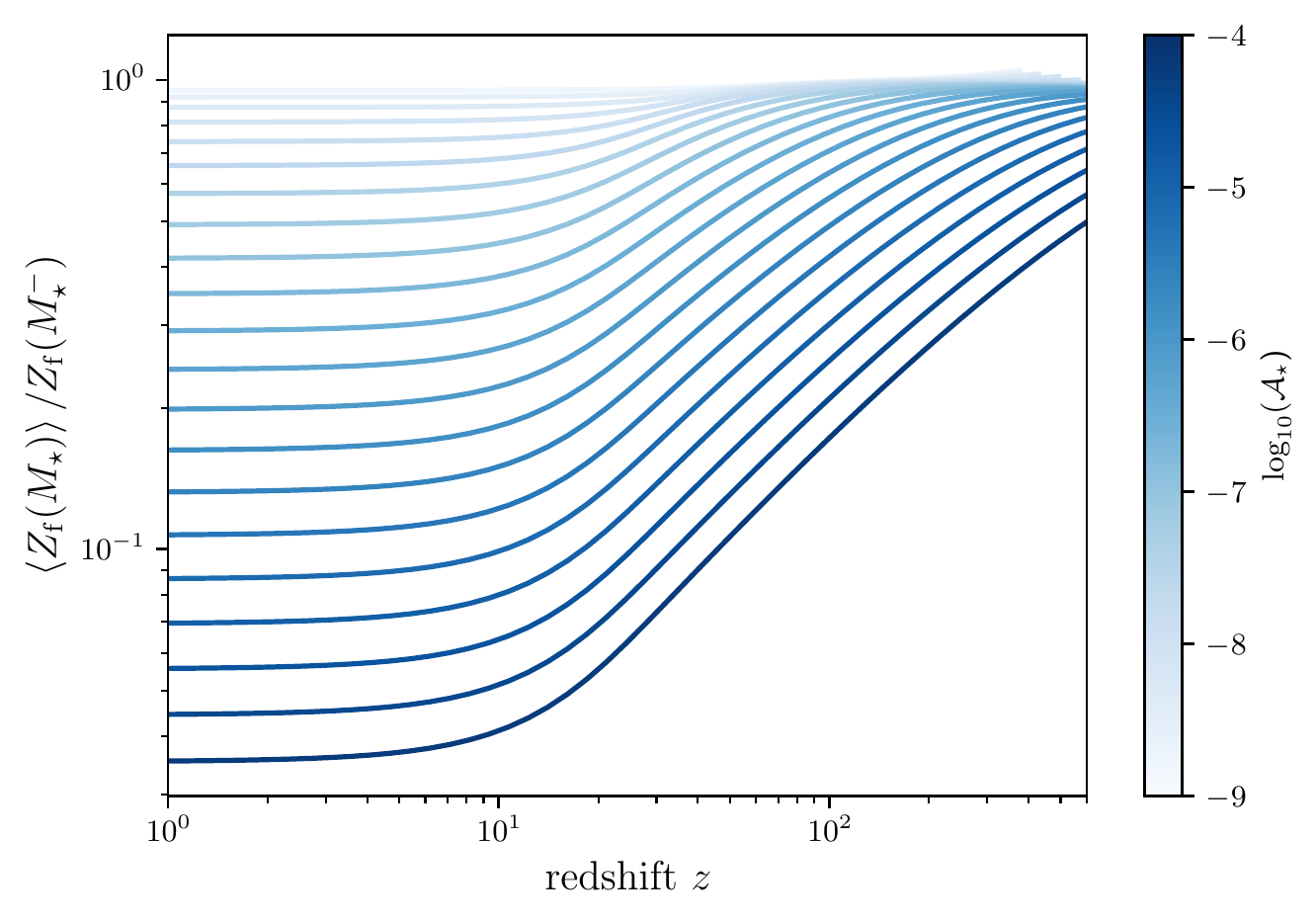}
\caption{ {\bf Left panel.} Average formation redshift  for halos \emph{from} the spike $\left< Z_{\rm f}(M_\star) \right>$ at $k_\star = 10^3$~$\rm Mpc^{-1}$ ($M_\star = 2.3 \times 10^3$ M$_\odot$). The blue color scale covers a range of amplitudes $\mathcal{A}_\star$, logarithmically distributed from $10^{-9}$ to $10^{-4}$. The red area spans the range of values where the averaged formation redshift is lower than the redshift (which is non physical). {\bf Right panel.} Similar figure than on the left panel but for the ratio of the average formation redshift over the upper bound $Z_{\rm f}(M_\star^-)$.}
\label{fig:Zf}
\end{figure}

Let us discuss now the definition of formation redshift with a spiky power spectrum. The value of the averaged formation redshifts for halos arising \emph{from} the spike, $\left< Z_{\rm f}(M_\star) \right>$, and its ratio with the upper bound $ Z_{\rm f}(M_\star^-)$ (see \refequ{eq:Zfaverage} and \refequ{eq:Zfaverage2}) are respectively shown in the left and right panel of \reffig{fig:Zf}. The collection of curves covers a range of amplitudes $\mathcal{A}_\star$ from $10^{-9}$ to $10^{-4}$ at fixed $k_\star = 10^3~{\rm Mpc^{-1}}$. The red area shows the unphysical region where the average formation redshift becomes lower than the redshift at which the halo is supposed to be observed. This effect occurs even when the mass fraction in the spike is non negligible (although neither overwhelmingly dominant), as seen in the left panel of \reffig{fig:VarianceMatterPowerSpectrum}. Nonetheless, in practice, as the formation redshift appears more as a mathematical tool to properly define the halo property we do not exclude this case. A lower formation redshift only results in less concentrated halos with a subsequent lower contribution to the boost -- \emph{c.f.}, \refequ{eq:dimensionlessOneHaloBoost}. An alternative solution would be to define the formation redshift from the much larger upper limit $ Z_{\rm f}(M_\star^-)$ indirectly shown in the right panel. However, the higher the amplitude the higher the average formation redshift and the larger the difference with the upper bound. Therefore, at high amplitude this choice would most probably overestimate the true formation redshift of the vast majority of the halos from the spike at low-$z$. In effect, notwithstanding that it does predict a much larger boost at late time as we illustrated in the left panel of \reffig{fig:boost_several_prescriptions} and discussed in \refsec{sec:boost_refined_model}, it has no effect at the epoch that matters the most for the impact of DM annihilation on the CMB anisotropies -- because the difference between $\left< Z_{\rm f}(M_\star) \right>$ and $ Z_{\rm f}(M_\star^-)$ gets lower at larger redshifts as shown in the left panel. In conclusion, we can thus safely consider the formation redshift of halos from the spike to be $\left< Z_{\rm f}(M_\star) \right>$. This choice may not appear as the most physical from the point of view of what could be called the redshift of the halo formation (depending on the chosen definition) by itself but it should be closer to a better description of reality when translating it in terms of an average halo concentration and contribution to the boost.

\section{Physical motivation for the prior range on $\As$}
\label{app:prior_region}

\begin{figure}[t!]
\centering
\includegraphics[width=0.8\linewidth]{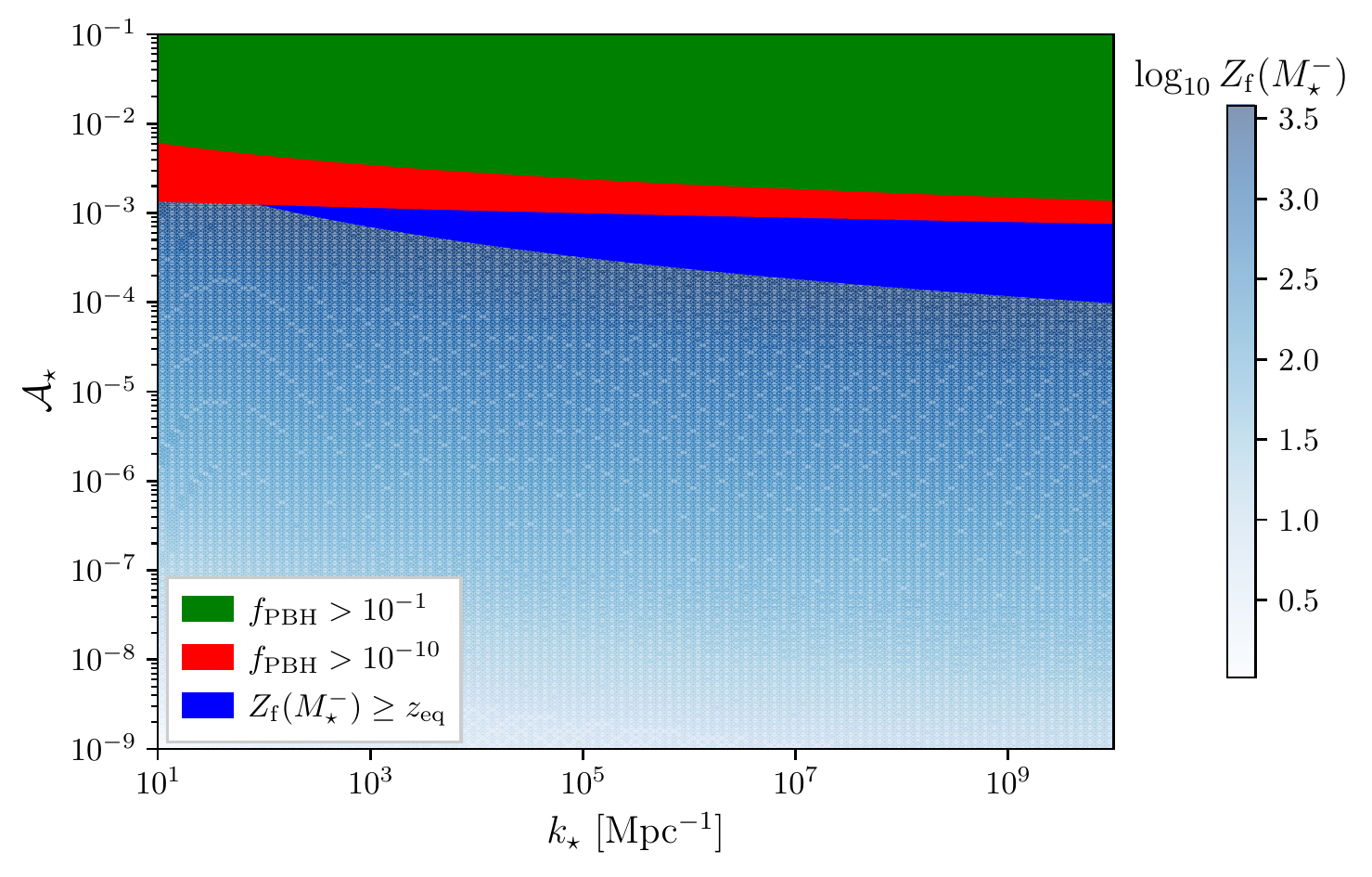} 
\caption{Fraction of DM in the form of PBH $f_{\rm PBH}$ due to a monochromatic enhancement of the power spectrum $\mathcal{P}_\star(k_\star) = \mathcal{A}_\star k_\star\delta(k - k_\star)$ and associated upper bound on the UCMHs formation redshift $Z_{\rm f}(M_\star^-)$. The value of the latter is represented by the blue color scale and the blue patch as soon as it gets larger than the redshift of matter-radiation equality. The fraction of PBH is shown by two areas where $f_{\rm PBH} > 10^{-10}$ (red) and $f_{\rm PBH} > 10^{-1}$ (green).}
\label{fig:constrain}
\end{figure}

In this appendix we detail the physical motivation for our chosen prior range on the spike amplitude $\mathcal{A}_\star$. Two requirements must be satisfied in order for our theoretical framework to hold:
\begin{itemize}
    \item as PBH formation can be triggered by a monochromatic enhancement of the power spectrum, we impose that the fraction of DM in the form of PBH $f_{\rm PBH}$ remains small.
    \item because our formalism implies that UCMHs form during the matter dominated era we limit ourselves to the region $Z_{\rm f}(M_\star^-) > z_{\rm eq} \simeq 3400$.
\end{itemize}
Both conditions are summarised in \reffig{fig:constrain}. We evaluate $f_{\rm PBH}$ following the derivation performed in section II.A. of Ref.~\cite{Green:2020jor}. The value of $f_{\rm PBH}$ is very sensitive to changes in $\mathcal{A}_\star$. Therefore any trouble that could arise from the presence of PBH is avoided by setting the condition $\mathcal{A}_\star < 10^{-3}$ (see \cite{Kohri:2014lza, Nakama:2019htb, StenDelos:2022jld} for mixed scenarios involving PBHs and minihalos). The upper bound for the formation redshift of UCMHs is evaluated according to our definition of $Z_{\rm f}(M_\star^-)$ in \refsec{sec:distrib_formation_redshift}. Stronger than the PBH bound, this second condition limits the parameter range at $\mathcal{A}_\star < 10^{-4}$. \\

As a matter of fact, we can derive an approximate expression for the latter limit. Let us start from the approximation of $\Sigma_\star$ in \refequ{eq:SigmaStarApprox} and notice that the limit is set at such large values of $\mathcal{A}_\star$ that $\Sigma_\star \gg S_{\rm PL}(R_\star)$. By definition of $Z_{\rm f}(M_\star^-)$ one has
\begin{equation}
    \sqrt{\Sigma_\star} = \frac{D(0) \delta_{\rm c}}{D(Z_{\rm f}(M_\star^-))}
\end{equation}
Since $D(z) \simeq 1/(1+z)$ in the matter dominated era and the limit is set for $Z_{\rm f}(M_\star^-) = z_{\rm eq}$, it yields that the upper bound is fixed by
\begin{equation}
     \sqrt{\Sigma_\star} \le  D(0) \delta_{\rm c} (1+z_{\rm eq}) \, .
\end{equation}
Plugging numbers yields
\begin{equation}
        \mathcal{A}_\star \lesssim 6.6 \times 10^{-2} \times \ln^{-2}\left(\frac{k_\star}{k_{\rm T}}\right) \, ,
\end{equation}
with $k_{\rm T} = 6.2\times 10^{-2}~{\rm Mpc^{-1}}$. We can easily check that the blue curve of \reffig{fig:constrain} follows this simple relation. At $k_\star = 10^{10}~{\rm Mpc^{-1}}$ the constraint is $\mathcal{A}_\star \lesssim 10^{-4}$, hence the chosen limit on the prior range.

\section{Impact of a finite free-streaming scale $k_{\mathrm{fs}}$ on the constraints}
\label{app:k_fs}

\begin{figure}[t!]
\centering
\includegraphics[width=0.75\linewidth]{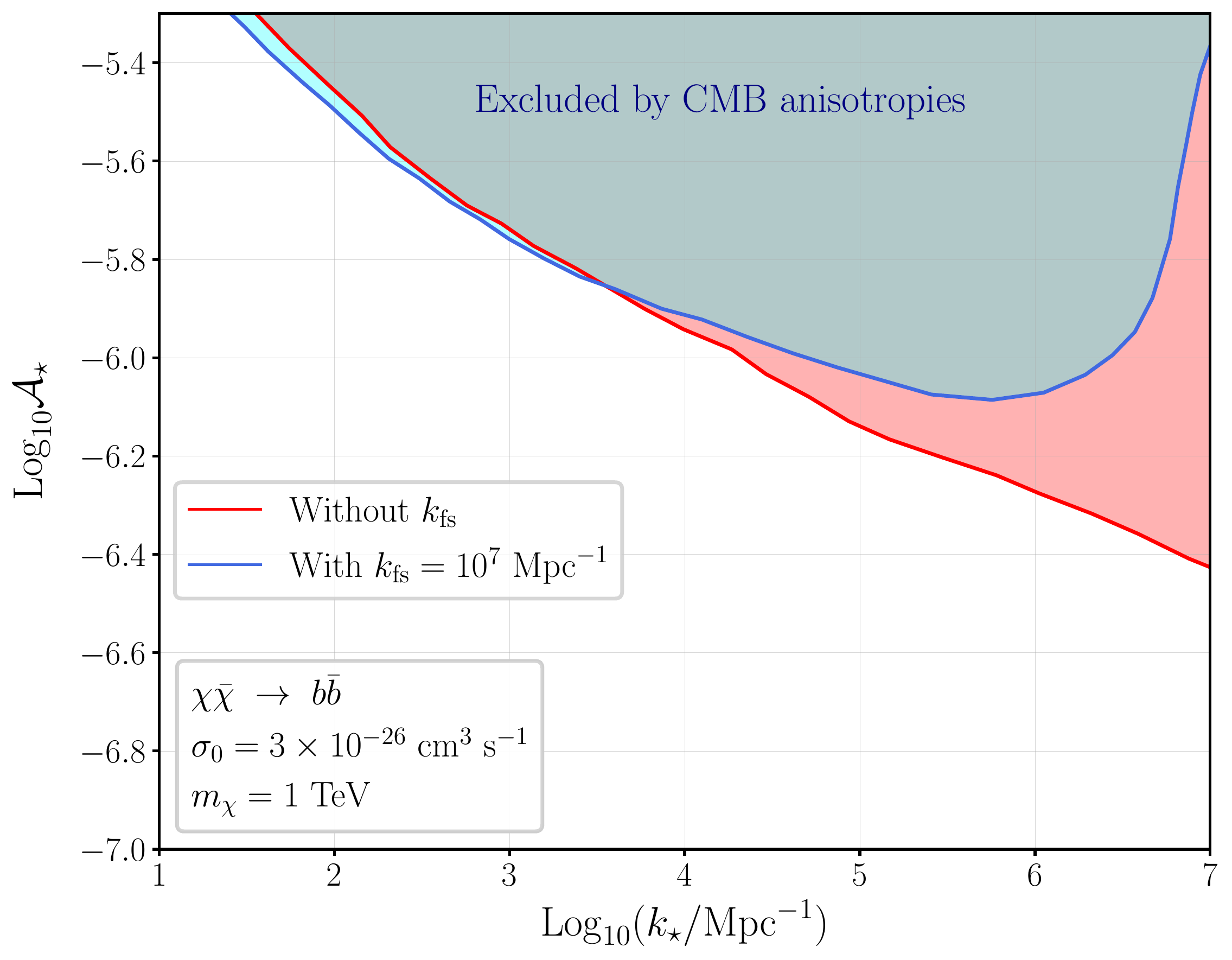} 
\caption{ 95\% C.L. bounds from \Planck+\BAO+\Pantheon~on the amplitude of the spike in the primordial spectrum as a function of its wavenumber, with and without the presence of a free-streaming scale $k_{\mathrm{fs}} = 10^7 \ \mathrm{Mpc}^{-1}$ in the transfer function. These curves were obtained with the EST formalism and  assuming that DM annihilates into $b\bar{b}$ with a mass $m_{\chi} =1 \ \rm{TeV}$ and a $s$-wave thermal relic cross-section.  }
\label{fig:constraints_wkfs}
\end{figure}

As already mentioned in \refsec{sec:results}, UCMH formation is expected to be suppressed when the velocity dispersion of typical WIMP particles starts to become relevant, i.e., at scales smaller than their free-streaming legth, $k_\star > k_{\rm fs}$. The value of $k_{\rm fs}$ is determined by the temperature of kinetic decoupling of WIMPs and therefore should be computed on a per-model basis, but it is generally of the order $k_{\rm fs} \sim 10^6 -10^7 \ \rm{Mpc}^{-1}$ \cite{Bringmann:2009vf}. For this reason, most of the  UCMH studies in the past have considered constraints in the primordial spectrum only up to $k_\star \sim  10^7 \ \rm{Mpc}^{-1}$\cite{Delos:2018ueo,Kawasaki:2021yek}. In practice, it is not entirely realistic to cut abruptly the constraints at $k_\star =k_{\rm fs}$, since UCMH formation is already partially suppressed at scales $k_\star \lesssim k_{\rm fs}$ and this can have some impact on the constraints. Motivated by this observation, we show in this appendix the effects of including a free-streaming scale $k_{\rm fs}$ in the calculation of the boost factor. This should reflect how the constraints on the amplitude of the spike $\As$ gradually relax as one gets closer and closer to $k_\star =k_{\rm fs}$. To account for this effect, we simply replace the transfer function in \refequ{eq:Pmatter} by $\tilde{T}(k)=T(k)D_{\rm fs}(k)$, where $D_{\rm fs}(k)$ is an exponential cutoff given by \cite{Green:2005fa}
\begin{equation}
D_{\rm fs}(k) = \left(1-\frac{2}{3} \left(\frac{k}{k_{\rm fs}}\right)^2 \right) \exp{\left[  -\left(\frac{k}{k_{\rm fs}}\right)^2\right] }.
\end{equation}
This leads to a steep suppression in the halo mass function for masses $M < M_{
\rm min} = \gamma_R k_{\rm fs}^{-3} \bar{\rho}_{\rm m, 0}$. In \reffig{fig:constraints_wkfs} we show the 95\% C.L. constraints on the $\As$ vs. $\ks$ plane with and without including the exponential cutoff at $\ks = 10^7 \ \mathrm{Mpc}^{-1}$. We see that both exclusion regions agree well for small values of $\ks$ (as they should), but start to deviate significantly at $ \ks \gtrsim 10^6 \ \mathrm{Mpc}^{-1}$. The sharp relaxation of the constraints close to the exponential cutoff is precisely the kind of physical behaviour that we were expecting to recover.


\addcontentsline{toc}{section}{Bibliography} 
\bibliographystyle{JHEP}
\bibliography{references_UCMH} 

\end{document}